\documentclass[aps,prl,showpacs,preprintnumbers,twocolumn,superscriptaddress]{revtex4-2}
\usepackage{amsmath,amssymb}
\usepackage{bm}
\usepackage{tipa}
\usepackage{upgreek}
\usepackage{comment}
\usepackage{mathrsfs}
\usepackage{graphicx}
\usepackage{braket}
\usepackage{mathbbol}
\usepackage{booktabs}
\usepackage{enumitem}
\usepackage{gensymb}
\usepackage[normalem]{ulem}
\usepackage{color}
\usepackage[colorlinks,bookmarks=true,citecolor=blue,linkcolor=red,urlcolor=blue]{hyperref}
\usepackage{hyperref}
\renewcommand{\vec}[1]{\mathbf{#1}}
\usepackage{pifont}

\newcommand{\xmark}{\ding{55}}
\newcommand{\bk}{\bm{k}}
\newcommand{\bG}{\bm{G}}

\newcommand{\bq}{\bm{q}}
\newcommand{\tk}{\tilde{k}}

\begin{document}

\title{Coulomb-driven band unflattening suppresses $K$-phonon pairing  in moir\'e graphene}
  \author{Glenn Wagner}
	\affiliation{Department of Physics, University of Zurich, Winterthurerstrasse 190, 8057 Zurich, Switzerland}
 \author{Yves H. Kwan}
	\affiliation{Princeton Center for Theoretical Science, Princeton University, Princeton NJ 08544, USA}
	\author{Nick Bultinck}
	\affiliation{Rudolf Peierls Centre for Theoretical Physics, Parks Road, Oxford, OX1 3PU, UK}
	\affiliation{Department of Physics, Ghent University, Krijgslaan 281, 9000 Gent, Belgium}
	\author{Steven H. Simon}
	\affiliation{Rudolf Peierls Centre for Theoretical Physics, Parks Road, Oxford, OX1 3PU, UK}
	\author{S.A. Parameswaran}
	\affiliation{Rudolf Peierls Centre for Theoretical Physics, Parks Road, Oxford, OX1 3PU, UK}

\begin{abstract}
It is a matter of current debate whether the gate-tunable superconductivity in twisted bilayer graphene is phonon-mediated or arises from  electron-electron interactions. The recent observation of the strong coupling of electrons to so-called $K$-phonon modes in angle-resolved photoemission spectroscopy experiments has resuscitated early proposals that $K$-phonons  drive superconductivity. We show that the bandwidth-enhancing effect of interactions drastically weakens both the intrinsic susceptibility towards pairing as well as the screening of Coulomb repulsion that is essential for the phonon attraction to dominate at low temperature. 
This rules out purely $K$-phonon-mediated superconductivity with the observed transition temperature of $\sim 1$~K. We conclude that  the unflattening of bands by Coulomb interactions challenges any purely phonon-driven pairing mechanism, and must be addressed by a successful theory of superconductivity in moir\'e graphene. 
\end{abstract}

\maketitle

\textit{Introduction.---} Superconductivity requires attractive interactions between electrons in order to form Cooper pairs. In conventional superconductors, the necessary attractive ``glue'' is provided by phonons, but pairing can arise in other ways, ranging from the exchange of collective excitations such as spin waves or plasmons, to `overscreening' of Coulomb interactions, as in the venerable Kohn-Luttinger mechanism \cite{KL}. A case in point is the cuprate high-temperature  superconductors: their phenomenology  is widely thought to be inconsistent with phonon-mediated pairing, leading to extensive efforts to explain the origin of the high transition temperatures ($T_c$) in these and related materials.

Since the discovery of gate-tunable superconductivity in twisted bilayer graphene (TBG) \cite{Cao2,SC1,SC2,SC3,SC4,SC5,Cao3}, intense debate has centered on the pairing mechanism and its similarity to that in the cuprates. The highest reported $T_c$ in TBG is around 3K \cite{Cao3}. This relatively high $T_c$ --- as compared to the small Fermi temperature  --- has motivated several theories of Coulomb-interaction-mediated superconductivity~\cite{cea2021coulomb,Khalaf,Sharma,Roy,Isobe,HUANG,Kennes,Po,christos2023nodal,Christos_2020,ChatterjeePRB,E1,E2,E3,E4,E5,E6,E7,E8,ingham2023quadratic,gonzalez2023universal},  whereas the myriad peculiarities of the moir\'e superlattice  structure  and electronic band topology have  stimulated a comparable number of proposals for phonon-mediated mechanisms~\cite{Lewandowski2021does,Lewandowski2021umklapp,Wu,liu2023electronkphonon,Lian,Wu_Hwang,Peltonen,Shavit,oppeneer2020prominent,christos2023nodal,Blason,ChoiPh,P1,P2,yu2022euler}.

Experimentally, there are tantalizing hints that phonons may drive superconductivity in TBG. For example, increasing the screening of the Coulomb interaction in TBG is shown to suppress the correlated insulators, while leaving $T_c$ relatively unchanged~\cite{Saito,Stepanov,Liu}. Furthermore, recent angle-resolved photoemission spectroscopy (ARPES) experiments have observed strong coupling of the electrons to a $K$-phonon mode \cite{chen2023strong}, and have noted that superconductivity is absent in devices where  coupling to this mode appears to be suppressed.

Two existing theoretical proposals for superconductivity 
consider $K$-phonon-mediated
pairing~\cite{Wu,liu2023electronkphonon}; however, the corresponding energy scale  ($\sim0.5$ meV) is  weak compared to the repulsive Coulomb interaction ($\sim20$ meV). The effectiveness of $K$-phonon pairing 
therefore relies on the screening of the Coulomb interaction due to the high density of states in the central bands of TBG. Theoretically incorporating screening at the Thomas-Fermi level and working with the narrow ($\sim1$ meV bandwidth) bands of the Bistritzer-MacDonald (BM) continuum model~\cite{Bistritzer2011}, Refs.~\onlinecite{Wu,liu2023electronkphonon} indeed find $K$-phonon-mediated superconductivity with $T_c\sim 1$K, that compares favorably with experiment.  

However,  scanning tunneling microscopy (STM) \cite{strain3,strain1,strain2,jiang}, compressibility \cite{comp} and ARPES \cite{expBW} experiments indicate that the bandwidth of the central bands is significantly larger ($\sim50$meV) than that of the bare  non-interacting BM model. This  
`band unflattening' is likely due to  both the interaction-induced renormalization of the band structure as well as the effect of strain, known to be important in TBG~\cite{Parker,IKS_PRL,IKS_PRX}. The increased bandwidth lowers the density of states, both weakening the pairing instability and suppressing screening. How does this 
impact
$K$-phonon superconductivity?

We answer this question by studying the pairing of electrons in the Hartree-Fock renormalized bands of TBG,
in the presence of Thomas-Fermi screened Coulomb interactions and $K$-phonon attraction. We show that in any realistic parameter regime, the enhancement of the bandwidth  and the concomitant suppression of the density of states 
leads to a significant reduction of 
$T_c$ to well below experimentally reported values. We find similar results irrespective of whether electrons are doped into the incommensurate Kekul\'e spiral state known to be the insulating ground state at non-zero integer filling and non-zero strain, the strong-coupling `ferromagnetic' insulators at zero strain, or the  
gapless (metastable) parent states of either phase: band unflattening is the key to suppressing $T_c$. Our results show that theories of phonon-mediated attraction in TBG must include screening  beyond the Thomas-Fermi approximation, or identify another route to evade the Coulomb suppression of pairing.

\textit{Hartree-Fock band structure}.--- The starting point for many studies of TBG are the BM bands, computed using the non-interacting continuum model of Ref.~\onlinecite{Bistritzer2011}. We also  begin here, choosing as BM parameters sublattice-dependent hopping matrix elements $w_\text{AB}=110\,$meV and $w_\text{AA}=80$meV, and a twist angle $\theta=1.1^\circ$, corresponding to a bare bandwidth $\sim1$ meV.
However, our next step is less commonly pursued: we consider the enhancement of the bandwidth by Coulomb interactions. We study this effect, expected to be significant, using the self-consistent Hartree-Fock (HF) mean-field approximation. (We note that HF has been successfully used to make a variety of nontrivial quantitative predictions in TBG \cite{dw,sk,IKS_PRX,IKS_PRL,GSHS,TBG3,TBG4,Hejazi,XieMD,Liu_HF,HFS,LiuDai,CEA202227,CeaHF}, and is expected to capture well the interaction-driven broadening of the bandwidth. Away from the flat-band regime, \cite{Lewandowski2021does,Choi} find an interaction-driven flattening of the bands, however even in that case the bandwidth is on the order of the interaction scale and the challenge to superconductivity described in our work remains \cite{SupMat}.) To match the experimental device setting, we assume a dual-gate screened Coulomb interaction $V^0(q)=e^2\tanh(qd)/(2\epsilon_0\epsilon_rq)$ with $\epsilon_r=10$ and gate distance $d=25$nm. We use the so-called `average' subtraction scheme to avoid double counting the interactions (see \cite{IKS_PRX} for details). 
To obtain the interaction-renormalized bandstructure, we perform a self-consistent HF calculation on a $10\times10$ grid for the moir\'e Brillouin zone (mBZ), and then interpolate the resulting bands to a $40\times40$ and $100\times100$ grid following the procedure described in Ref.~\cite{TSTG2}. The latter step is done in order to obtain a more finely resolved band structure on which to compute the density of states ($100\times100$ grid) and to solve the superconducting gap equation ($40\times40$ grid).

\textit{$K$-phonons, screening, and the gap equation}.--- TBG hosts a variety of phonon modes; both acoustic phonons and phasons experience strong moir\'e effects --- indeed, the latter result from the incommensurate nature of the moir\'e superlattice --- and  both have been proposed as drivers of  superconductivity.  However, there are other graphene phonon  modes that persist with relatively little moir\'e reconstruction in TBG, yet are very strongly coupled to the moir\'e bands. Motivated by the recent observation of a particularly strong coupling of the central-band electrons to the $A_1$ zone-corner   optical phonon mode  \cite{chen2023strong}, we focus here on pairing due to this `$K$-phonon' mode. $K$-phonon coupling can favour different insulating states \cite{Blason} and in Ref.~\cite{kwan2023electronphonon} it has  been mooted as a route to stabilizing the Kekul\'e charge-order seen in  STM  on ultra-low strain samples~\cite{nuckolls2023quantum}. Accordingly, we consider the phonon-induced electron-electron interaction due to $A_1$ phonons at momentum $K_\text{D}$,
\begin{equation}
  \negthinspace\negthinspace \negthinspace H_\text{ph}=-g\sum_l \negthinspace\int \negthinspace d^2r\left[\left(\psi^\dagger_l\tau_x\sigma_x\psi_l\right)^2
+\left(\psi^\dagger_l\tau_y\sigma_x\psi_l\right)^2\right] \negthinspace,
\end{equation}
where 
$\psi$ is a spinor in sublattice space, $l$ is the layer index, $\sigma_i$ ($\tau_i$) are Pauli matrices in sublattice (valley) space, and  $g$ is the coupling constant $\simeq 69 \,\text{meVnm}^2$ \cite{Wu}. As noted above, two previous studies have considered $K$-phonon mediated superconductivity: Ref.~\cite{Wu} finds a maximum $T_c$ of around 10K in the non-interacting problem; Ref.~\cite{liu2023electronkphonon} incorporates Thomas-Fermi screened Coulomb repulsion and finds that this reduces $T_c$  to around 2K. These results appear consistent with the experimental $T_c$ of around 1-3K. However, crucially --- as already pointed out in Ref.~\cite{liu2023electronkphonon} --- this is likely an overestimate of $T_c$ since both these works solve the BCS gap equation for the BM model at the magic angle, which has an extremely small bandwidth. The interaction scale in TBG is large compared to the bandwidth; as we show below, accounting for the former via Hartree-Fock yields  a significant enhancement of the latter, consistent with experimental measurements and quantum Monte Carlo studies~\cite{qmc}. Both suggest that the actual bandwidth is least an order of magnitude larger than that of the magic-angle BM model. Therefore, instead of starting from the BM bandstructure, a more accurate approach is to solve the gap equation for the HF bandstructure as we now do.  Besides the phonon-mediated attraction, we can also incorporate screened repulsive Coulomb interactions, 
\begin{equation}
    V(q)=\frac{V^0(q)}{1-\Pi(q)V^0(q)}
\end{equation}
where we make the Thomas-Fermi approximation of a constant polarization function $\Pi(q)\approx -D(E_F)$. As we note in the discussion below, it is in principle possible though technically much more demanding to account for screening in a more refined manner, e.g. by computing the polarization bubble in the random-phase approximation (RPA). This may allow effects such as over-screening which may be able to seed superconductivity. However, the numerical challenge in implementing these  extensions renders them beyond the scope of the present work.

For a given band structure,  we solve the gap equation
\begin{equation}
    \Delta_{ab}(k)=-\frac{1}{A}\sum_{k',cd}U_{abcd}(k,k')\pi_{cd}(k')\Delta_{cd}(k'),
\end{equation}
where roman letters indicate a band label, $U_{abcd}(k,k')$ is the BCS-channel vertex --- either a purely phonon attraction or contains both phonon and Coulomb contribution --- and $\pi_{cd}(k')$ is the particle-particle susceptibility. At low temperatures, $\pi$ is sharply peaked at the Fermi surface. We therefore evaluate the momentum sum on a set of momentum points obtained via importance sampling from the susceptibility~\cite{SupMat}. We solve the gap equation at a fixed temperature $T$: recall that as $T_c$ is approached from above, the leading eigenvalue $\lambda(T_c)$ drops below $-1$, so that $\lambda(T)\leq -1$ indicates a superconducting solution.

\begin{figure*}
    \centering
    \includegraphics[width=\textwidth]{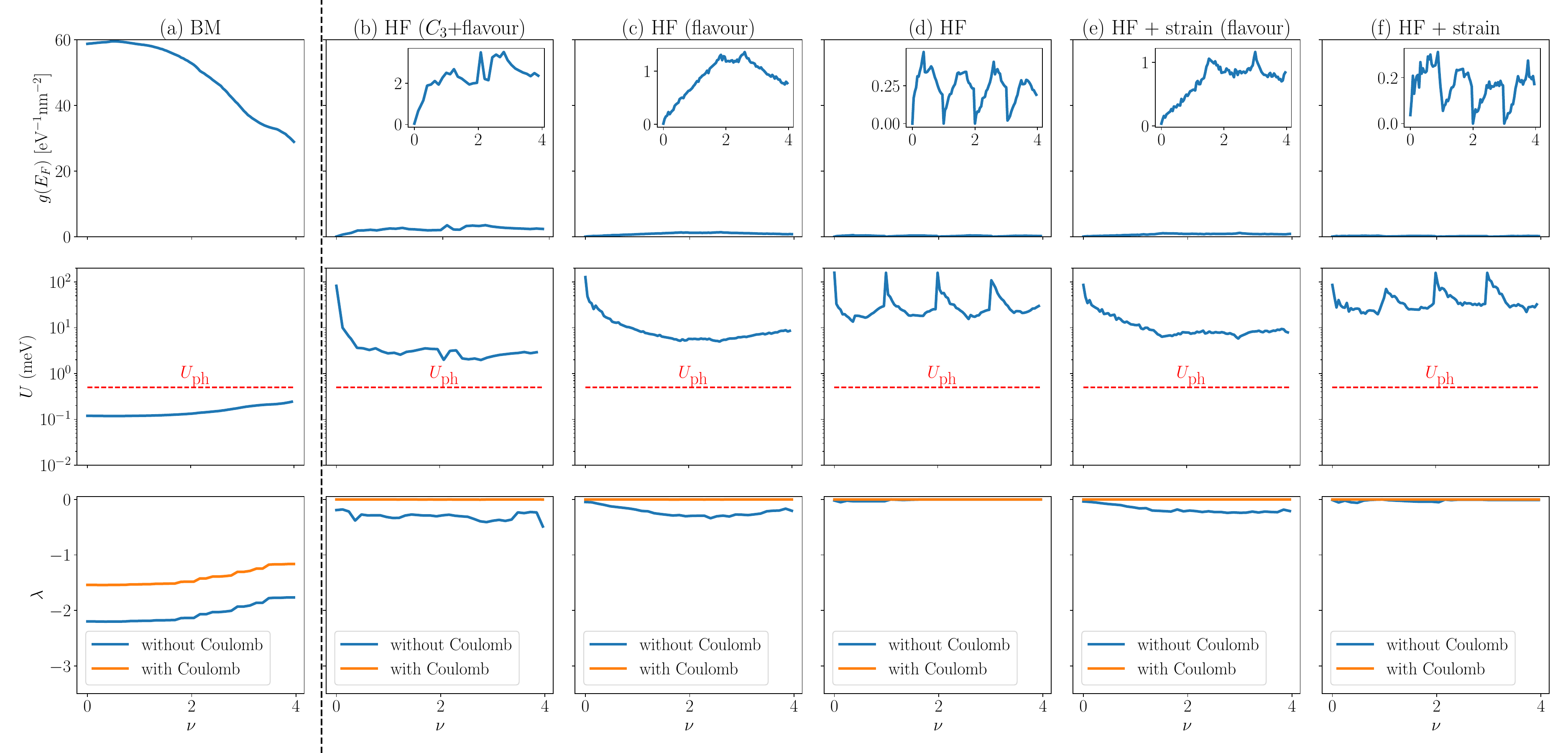}
    \caption{\textbf{Suppression of pairing by Coulomb interactions}. Column (a) shows  BM model results as a reference; columns (b)-(d) show  results for unstrained HF-renormalized bands that respectively preserve both flavour and $C_3$ symmetry, only flavour, and break both; (e) and (f) show results for strained HF-renormalized bands with and without flavour symmetry. (cf. Tab.~\ref{tab:bandstructures}).
    {\bf Top row:} Density of states at the Fermi level with a broadening corresponding to temperature $T=2\,$K.%$T=2\,$K.
    {\bf Middle row:} Energy scales of the Thomas-Fermi screened Coulomb interaction $U=V(q=0)$, 
    and of the phonon-mediated attraction  $U_\textrm{ph}=\frac{g}{A_M}=$0.5meV, where $A_M$ is the moir\'e unit cell area. Even for the highest level of screening, $U\gg U_\textrm{ph}$,  suppressing superconductivity.  {\bf Bottom row:} Gap equation eigenvalue $\lambda$  without and with Coulomb interactions at temperature $T=2\,$K, where $\lambda=-1$ signifies a superconducting instability. At HF level, interactions reduce $\lambda$ by more than an order of magnitude, reducing $T_c$ well below the experimentally observed scale.}
    \label{fig:all_band_results}
\end{figure*}

\begin{table}[b!]
\centering
\newcommand{\colskip}{\hskip 0.12in}
\renewcommand{\arraystretch}{1.1}
\begin{tabular}{
l @{\hskip 0.05in} 
c @{\colskip}  
c @{\colskip} 
c @{\colskip} 
c @{\colskip} 
c @{\colskip} 
c  }\toprule[1.3pt]\addlinespace[0.3em]
Bandstructure  & strain &
$C_3$ & 
$SU(2)$ & 
$U_V(1)$ & 
$\mathcal{T}$ \\ (symmetries enforced) & & & & &
\\ \midrule
(a) BM & - &  \checkmark & \checkmark & \checkmark & \checkmark\\ %\hdashline
(b) HF ($C_3$+flavour)& - &  \checkmark & \checkmark & \checkmark & \checkmark\\
(c) HF (flavour)& -&  \xmark (S) & \checkmark & \checkmark & \checkmark\\
(d) HF& - &  \xmark (S) & \xmark & \xmark & \xmark\\
(e) HF (flavour)& 0.3\%  & \xmark (E) & \checkmark & \checkmark & \checkmark\\
(f) HF & 0.3\% &  \xmark (E) & \xmark & \xmark & \checkmark
\\\bottomrule[1.3pt]
\end{tabular}
\caption{\label{tab:bandstructures}\textbf{Six bandstructures investigated.} We enumerate the six bandstructures that we considered as the parent state for superconductivity. The different bandstructures have different symmetries enforced (the symmetries enforced being a subset of $C_3$ rotation, spin $SU(2)$, valley $U_V(1)$ and time-reversal $\mathcal{T}$). We list the symmetries which are either broken (\xmark) or preserved (\checkmark) by each bandstructure. In the case of $C_3$ we also list whether the symmetry is broken spontaneously (S) or explicitly (E).}
\end{table}

\textit{Results.---} The main results of this work are summarized in Fig.~\ref{fig:all_band_results}, that shows the outcome of solving the gap equation starting with the HF-renormalized bands of several different physically-motivated choices of integer-filling parent state, listed in Table~\ref{tab:bandstructures} and described below. Each column of Fig.~\ref{fig:all_band_results} corresponds to a distinct choice of input bandstructure to the gap equation. As a function of doping, the top row shows the density of states, the middle row the energy scales of the $K$-phonon attraction (red, dashed) and the Thomas-Fermi screened Coulomb interaction (blue, solid), and the bottom row the results of solving the gap equation with (orange) and without (blue) including the screened Coulomb repulsion. 
In each case, we work at fixed $T= 2\,$K, consistent with the typical $T_c$ seen in experiment, and focus on positive fillings $\nu$ (i.e.~the number of electrons per moir\'e unit cell relative to charge neutrality) due to the approximate particle-hole symmetry of the model.

First, as a benchmark, Fig.~\ref{fig:all_band_results}(a) shows results for the non-interacting BM model. The sharp features of the BM bands are smeared out by the temperature, which is a larger energy scale than the bare bandwidth. For $T=2\,$K $\lambda(T)<-1$ and hence there is superconductivity at all fillings $\nu$ even in the presence of the Coulomb interaction, consistent with previous computations on this model~\cite{Wu,liu2023electronkphonon}. In the Supplement \cite{SupMat} we show that for all fillings $\lambda(T)>-1$ at $T=5$K, hence the (filling-dependent) critical temperature for the BM bands lies between 2K and 5K for our choice of paramters. 

Fig.~\ref{fig:all_band_results}(b-d) show the results of incorporating HF renormalization for unstrained TBG with differing degrees of symmetry breaking. In Fig.~\ref{fig:all_band_results}(b), we perform HF on the bands, resulting in a broadened bandwidth, but preserve the full symmetries of the BM model. In Fig.~\ref{fig:all_band_results}(c), we allow the breaking of $C_3$ symmetry but preserve flavour symmetry. At integer $\nu$, this allows us to access a metastable nematic (semi)metal \footnote{This state is semimetallic at $\nu=0$, while it is metallic at all other $\nu$.}, while suppressing  flavour-symmetry-broken correlated insulators.  This (semi)metal is the closest competitor to the strong-coupling insulators (such as the KIVC state) at zero strain (and becomes the ground state at $\nu=0$ for modest strains comparable to those seen in experiments~\cite{strain1,strain2,strain3}, see below). We therefore expect that it is the relevant gapless parent state for SC when the correlated insulators are suppressed by strain~\cite{Parker} or disorder~\cite{KIVCAnderson}; its open Fermi surfaces~\cite{SupMat} are consistent with magnetotransport data reported in Ref.~\cite{wang2022unusual}. In Fig.~\ref{fig:all_band_results}(d), we allow both $C_3$ and flavour symmetries to break spontaneously,  allowing gapped `generalized ferromagnets' at  integer $\nu$.

 Finally, Fig.~\ref{fig:all_band_results}(e,f) show results for the HF-renormalized bands of TBG subject to   $0.3\%$ heterostrain (which explicitly breaks $C_3$ and is  known to stabilize a semimetallic state at $\nu=0$ \cite{Parker}) both with and without  flavour symmetry. When flavour symmetry is preserved, the ground state at non-zero integer $\nu$ remains gapless, but on allowing flavour symmetry breaking it is the `incommensurate Kekulé spiral' state, which is gapped at $\nu=2,3$ and gapless at $\nu=1$~\cite{IKS_PRX,IKS_PRL}.

 Despite differences in details, the bandwidth in each case (b-f) is significantly broadened relative to the bare BM model and is typically on the order of 30~meV. (Quantum Monte Carlo at charge neutrality for the unstrained model~\cite{qmc} finds a similar bandwidth, but suffers from a sign  problem for $\nu\neq0$. An RG study also finds a renormalization of the bandwidth by the interactions \cite{Vafek}.)
 The enhanced bandwidth results in a  suppression of the density of states by nearly two orders of magnitude in  Fig.~\ref{fig:all_band_results}(b-f), which is reflected in a significantly stronger Coulomb repulsion due to the lowered screening. The two effects --- suppression of the density of states and reduced screening --- are both strongly antagonistic to $K$-phonon superconductivity, as reflected by the dramatic change in the gap  eigenvalue on their inclusion: a $2\,$K SC is ruled out in all cases (b)-(f).  Given the exponential dependence of the gap eigenvalue $\lambda$ on $T$, these results indicate that phonon-mediated/Thomas-Fermi screened SC mechanisms would yield $T_c$'s well below the experimentally observed values.  This conclusion holds irrespective of the choice of parent state, suggesting that the key mechanism that suppresses SC is the Coulomb-driven band unflattening rather than competing orders or detailed structure of the dispersions.

Each choice of parent state yields a distinct sequence of Lifshitz transitions and van Hove singularities (detailed in the Supplement~\cite{SupMat}). 
These features, as expected, lead to doping-dependent modulations of the gap eigenvalue, but on a scale much lower than the overall suppression due to enhanced bandwidth. Hence they  do not give a plausible route to $K$-phonon-driven SC near special van Hove fillings at temperature scales relevant to experiment.

\begin{figure}
    \centering
    \includegraphics[width=\columnwidth]{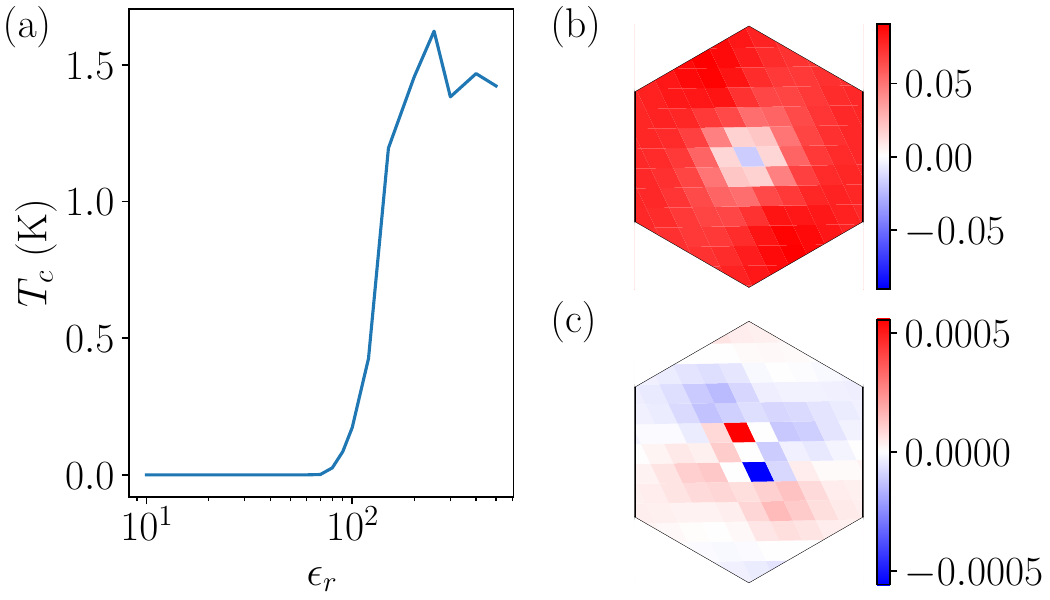}
    \caption{\textbf{Interaction-dependence of the critical temperature and symmetry of the gap function}. (a) As the Coulomb interaction is reduced by increasing $\epsilon_r$, $T_c$ increases to around 2K. $T_c$ only reaches the experimental value of around 2K for an unrealistically weak Coulomb interaction with $\epsilon_r>100$. (b,c)  mBZ-even and mBZ-odd components of the gap function. All results are for filling $\nu=2.48$.}
    \label{fig:changingepsilon}
\end{figure}

One route to stabilize a $K$-phonon-mediated SC is to suppress the Coulomb scale by screening, which can be adjusted by changing the distance to a metallic back gate. Fig.~\ref{fig:changingepsilon} shows more detailed results at fixed filling $\nu=2.48$ (close to the peak of the superconducting dome seen in many experiments) for the $C_3$-breaking, flavour-symmetric state  studied in Fig.~\ref{fig:all_band_results}(c). Fig.~\ref{fig:changingepsilon}(a) shows the effect of tuning interactions via  
the relative permittivity $\epsilon_r$, which enters the calculation in two ways. First, it controls the Hartree-Fock renormalization of the bandstructure, which reduces to the noninteracting BM bands as $\epsilon_r\to\infty$. 
Second, $\epsilon_r$ appears directly in the Coulomb interaction contribution to the BCS vertex in the gap equation: for $\epsilon_r\to\infty$ we recover a pure phonon-only attractive vertex. Both these effects tend to increase $T_c$ when increasing $\epsilon_r$.  $T_c$ is comparable to 2K once $\epsilon_r>100$. This is roughly consistent with a back-of-the envelope estimate of the effect of band renormalization on the results of  Ref.~\cite{liu2023electronkphonon}, and Fig.~\ref{fig:all_band_results}(a) which find $T_c\sim2\,$K for the {\it free} BM bands and screened Coulomb interactions. However this is an unrealistically large value, as we would typically expect  $\epsilon_r\sim 10$. Nevertheless, for completeness in Fig.~\ref{fig:changingepsilon}b-c we show the 
(b) mBZ-even and (c) mBZ-odd components of the leading gap function. Due to  fermionic anticommutation and the non-trivial action of inversion, that exchanges the valley degree of freedom ($\mathcal{I}\propto\tau_x$),
the gap structure is quite rich. An even-parity  gap function (which is forced to be a spin singlet) can have both mBZ-even, valley-triplet and mBZ-odd, valley-singlet components. Similarly, an odd-parity (hence  
spin-triplet) gap function  can involve both  mBZ-odd, valley-triplet and mBZ-even, valley-singlet contributions~\cite{SupMat}.

\textit{Discussion}.--- Upon discovery of superconductivity in TBG, immediate comparisons with the high-temperature superconductors were drawn. In particular, the emergence of a superconducting dome upon doping a correlated insulator is reminiscent of the phase diagram of the cuprates. However, recent work has shown that the correlated insulator is a symmetry-broken state quite unlike the Mott insulator in the cuprates  \cite{nuckolls2023quantum,IKS_PRX}. Despite this, $T_c$ in TBG is high compared to the Fermi temperature $T_F$ ($T_c/T_F\sim0.1$ \cite{Cao2}) and therefore the mechanism behind superconductivity remains a matter of debate. STM measurements support a nodal gap~\cite{SC5}, which requires non-$s$-wave pairing (or some topological obstruction \cite{yu2022euler}),  often viewed as a hallmark of unconventional superconductivity. 
Despite this, proposals of BCS-like phonon-mediated superconductivity abound in the literature, typically resolving the apparent conflict between a BCS-like mechanism and large  $T_c/T_F$ by solving the gap equation for the narrow-bandwidth magic-angle BM model. The high density of states in the flat bands compensates for the  the very low density of free electrons in doped TBG, leading to $T_c$'s comparable to experiment. However,  it is likely that many of these studies substantially underestimate the bandwidth: there is both theoretical and experimental evidence that effects such as strain and interaction renormalization introduce substantial dispersion so that the bands are not really flat. We show that for such interaction-broadened bands, a superconducting calculation based on pairing from a specific optical phonon mode (`$K$-phonon’) leads to a $T_c$ far below the experimental values. Other purely phonon-mediated mechanisms likely suffer similar problems. 

At minimum, future studies of superconductivity  should incorporate Coulomb interactions beyond the simple Thomas-Fermi approximation employed here. Given that the interactions set the largest energy scale in the problem, it is clearly important to treat them carefully. For example, already at the RPA level it is possible for  Coulomb interactions to develop an attractive component \cite{cea2021coulomb,E5,gonzalez2023universal,Goodwin}. The question of mechanism then rests on  the relative contributions of Coulomb interactions and phonons to the pairing. Other possibilities include a more intricate interplay of electron-electron   and electron-phonon interactions, as captured for example by the ``heavy fermion'' picture of TBG \cite{Heavy_fermion}.

Finally, we flag the insights provided by experiments that observe a relatively modest suppression of $T_c$ on increasing screening, but a dramatic suppression of the correlated insulators. Purely phonon-mediated attraction should show a significant  $T_c$ enhancement as the Coulomb repulsion is suppressed (as we show in Fig.~\ref{fig:changingepsilon}(a)). In contrast, the naive expectation for purely Coulomb-driven pairing  is that $T_c$ will decrease as screening increases. Thus, we conclude that the situation is likely more complicated, in that $T_c$ may not depend monotonically on the interaction strength. In general, we anticipate that the attractive interaction may involve  contributions from both phonons and Coulomb interactions, and it is quite likely that these are comparable in TBG.

\begin{acknowledgements}
\textit{Acknowledgements}.---We thank B.A.~Bernevig for comments on a previous version of this manuscript. This project was
supported by the European Research Council (ERC) under the European Union’s Horizon 2020 research and innovation program grant nos.~ERC-StG-Neupert-757867-PARATOP (GW) and ERC-StG-Parameswaran-804213-TMCS (SAP) and by EPSRC grant EP/S020527/1 (SHS). GW also acknowledges funding from the University of Zurich postdoc grant FK-23-134.

\end{acknowledgements}

\bibliography{refs.bib}

\newpage
\clearpage

\begin{appendix}
\onecolumngrid
	\begin{center}
		\textbf{\large --- Supplementary Material ---\\ Coulomb-driven band unflattening suppresses $K$-phonon pairing  in moir\'e graphene}\\
		\medskip
		\text{Glenn Wagner, Yves H. Kwan, Nick Bultinck, Steven H. Simon and S.A.~Parameswaran}
	\end{center}
	
		\setcounter{equation}{0}
	\setcounter{figure}{0}
	\setcounter{table}{0}
	\setcounter{page}{1}
	\makeatletter
	\renewcommand{\theequation}{S\arabic{equation}}
	\renewcommand{\thefigure}{S\arabic{figure}}
	\renewcommand{\bibnumfmt}[1]{[S#1]}

\section{Results for different interaction-renormalized bandstructures}

\begin{figure*}[h]
    \centering
    \includegraphics[width=\textwidth]{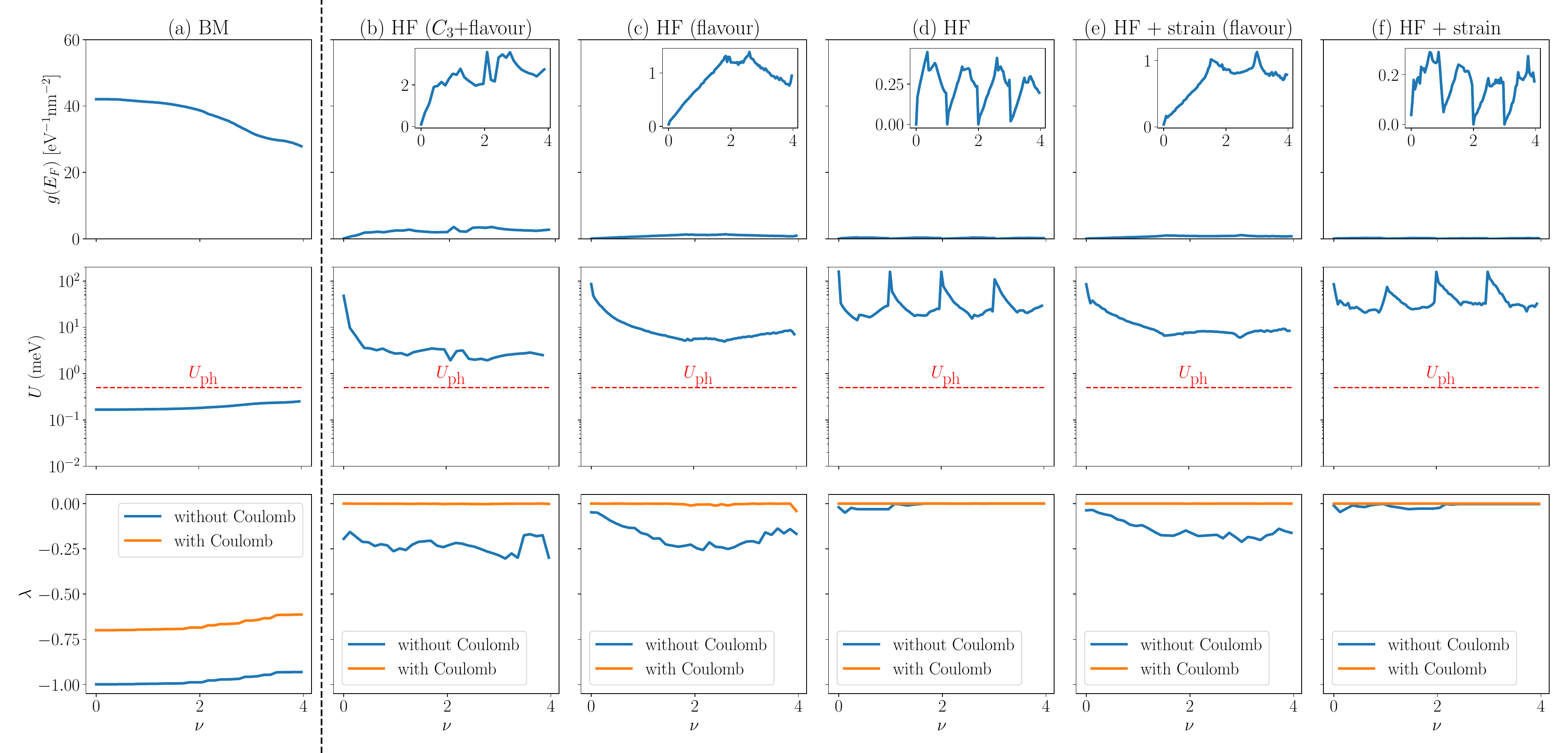}
    \caption{Same as Fig. 1 from the main text but at a temperature of 5K.
    {\bf Top row:} Density of states at the Fermi level with a broadening corresponding to temperature $T=5\,$K.
    {\bf Middle row:} Energy scales of the Thomas-Fermi screened Coulomb interaction $U=V(q=0)$, 
    and of the phonon-mediated attraction  $U_\textrm{ph}=\frac{g}{A_M}=$0.5meV, where $A_M$ is the moir\'e unit cell area. {\bf Bottom row:} Gap equation eigenvalue $\lambda$  without and with Coulomb interactions at temperature $T=2\,$K, where $\lambda=-1$ signifies a superconducting instability.}
    \label{fig:all_band_results_5K}
\end{figure*}

We discuss here in more detail the six bandstructures investigated in the main text:
\begin{enumerate}[label=(\alph*)]
    \item unstrained BM model 
    \item unstrained HF, $C_3$, valley $U(1)$ and spin $SU(2)$ symmetry enforced 
    \item unstrained HF, valley $U(1)$ and spin $SU(2)$ symmetry enforced
    \item unstrained HF
    \item 0.3\% strained HF, valley $U(1)$ and spin $SU(2)$ symmetry enforced
    \item 0.3\% strained HF. 
\end{enumerate}
In Fig.~\ref{fig:all_band_results_5K} we show the same result as in the main text but at $T=5$K. For the BM bands, there is superconductivity at all $\nu$ at $T=2$K, even in the presence of Coulomb repulsion. For $T=5$K the superconductivity is suppressed, showing that the critical temperature lies between 2K and 5K for all fillings $\nu$. For all HF bandstructures, the bandwidth is similar (since it is set by the Coulomb interaction strength, which is similar in all cases) and there is no superconductivity at $T=2$K in any of these cases.

\begin{figure}[h]
    \centering
    \includegraphics[width=0.7\textwidth]{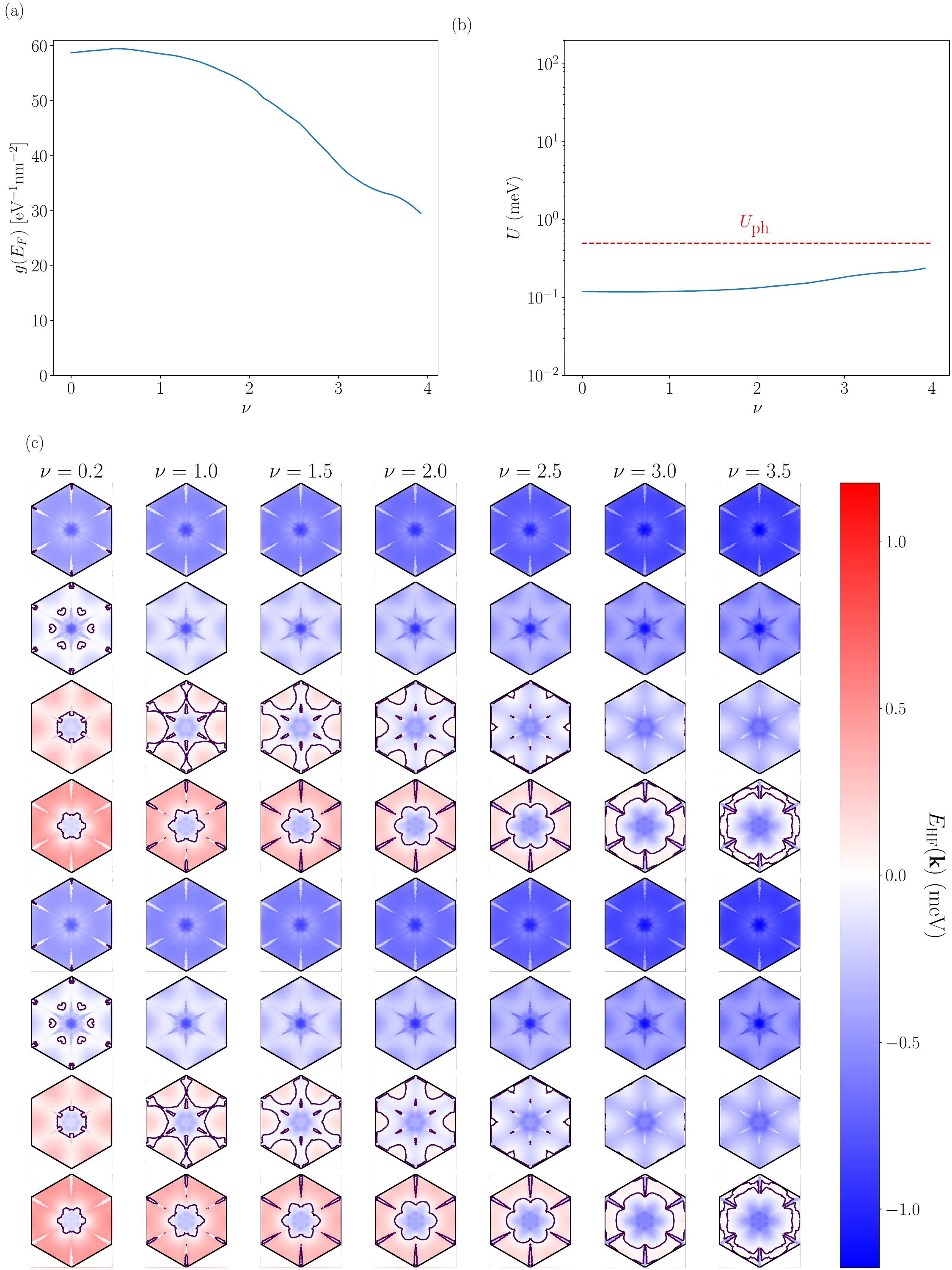}
    \caption{\textbf{Bandstructure (a).} (a) Density of states at the Fermi level with a temperature broadening corresponding to 2K. (b) Coulomb interaction scale $U$ compared to the phonon interaction scale $U_\textrm{ph}$. (c) Bandstructure and Fermi surface for seven representative fillings $\nu$. At each filling we plot the eight different bands (first four: up spin bands, last four: down spin bands). }
    \label{fig:SM1_0}
\end{figure}

\begin{figure}[h]
    \centering
    \includegraphics[width=0.7\textwidth]{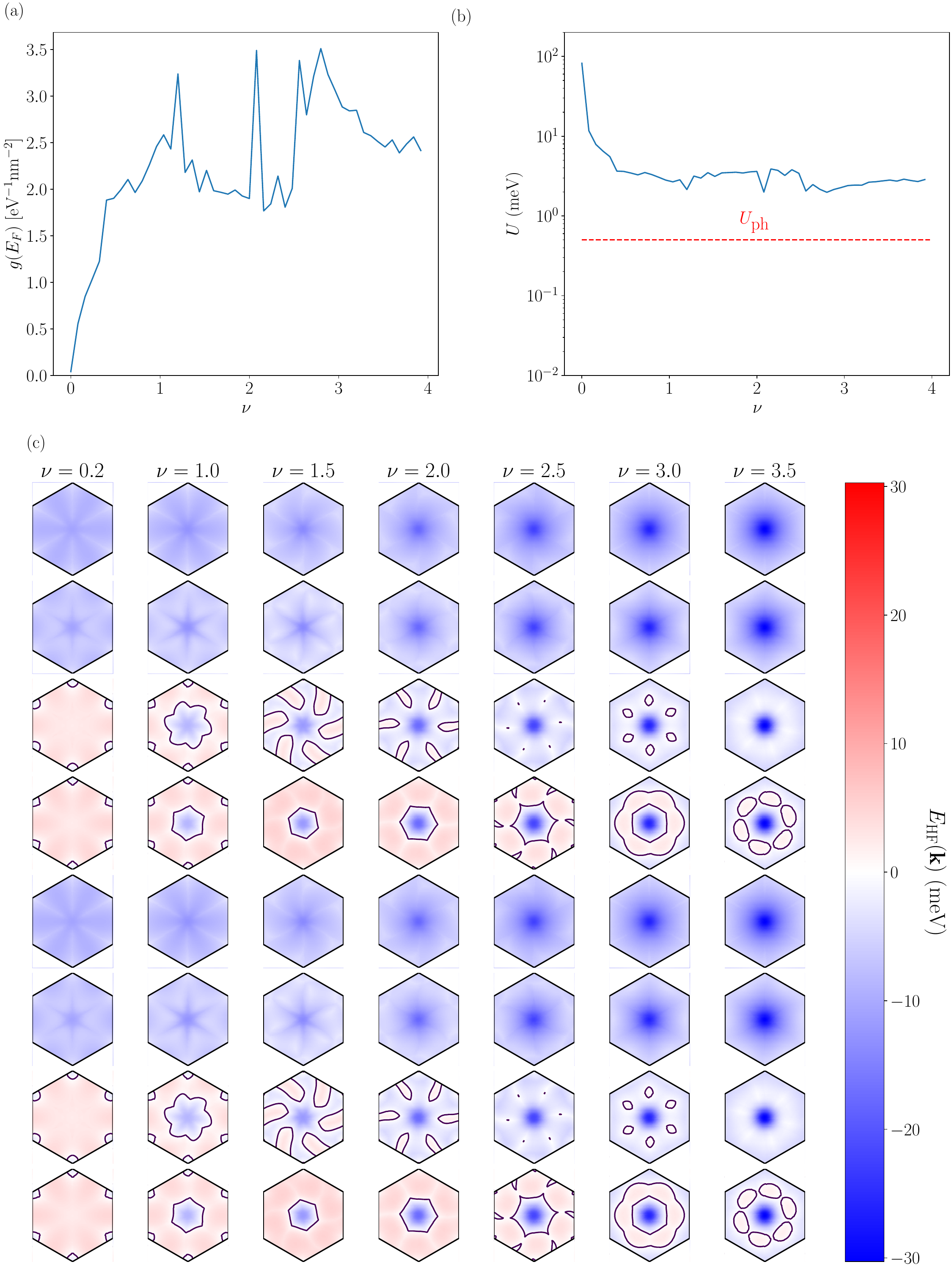}
    \caption{\textbf{Bandstructure (b).} (a) Density of states at the Fermi level with a temperature broadening corresponding to 2K. (b) Coulomb interaction scale $U$ compared to the phonon interaction scale $U_\textrm{ph}$. (c) Bandstructure and Fermi surface for seven representative fillings $\nu$. At each filling we plot the eight different bands (first four: up spin bands, last four: down spin bands). }
    \label{fig:SM1_1}
\end{figure}

\begin{figure}[h]
    \centering
    \includegraphics[width=0.7\textwidth]{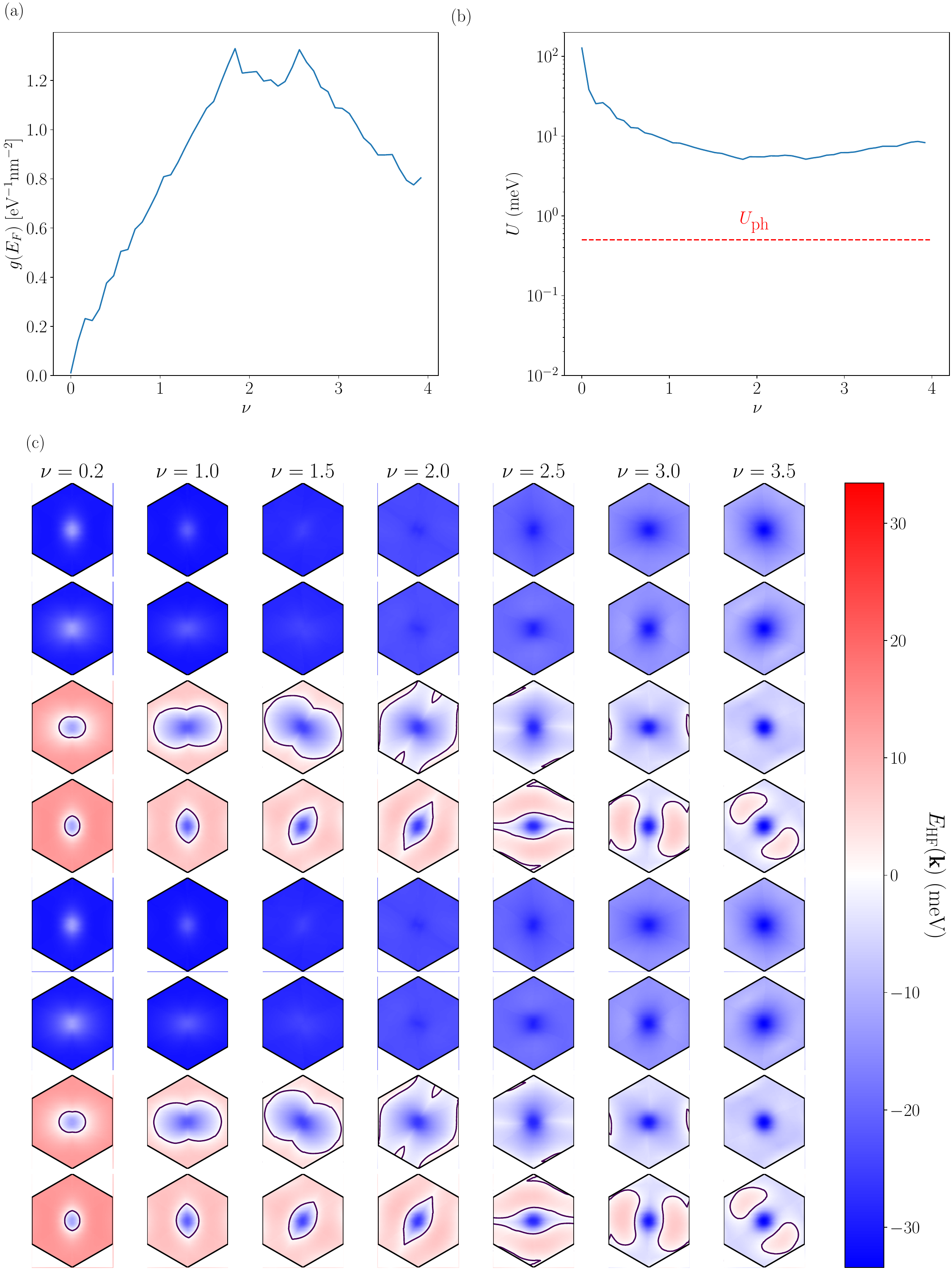}
    \caption{\textbf{Bandstructure (c).} (a) Density of states at the Fermi level with a temperature broadening corresponding to 2K. (b) Coulomb interaction scale $U$ compared to the phonon interaction scale $U_\textrm{ph}$. (c) Bandstructure and Fermi surface for seven representative fillings $\nu$. At each filling we plot the eight different bands (first four: up spin bands, last four: down spin bands). }
    \label{fig:SM1_2}
\end{figure}

\begin{figure}[h]
    \centering
    \includegraphics[width=0.7\textwidth]{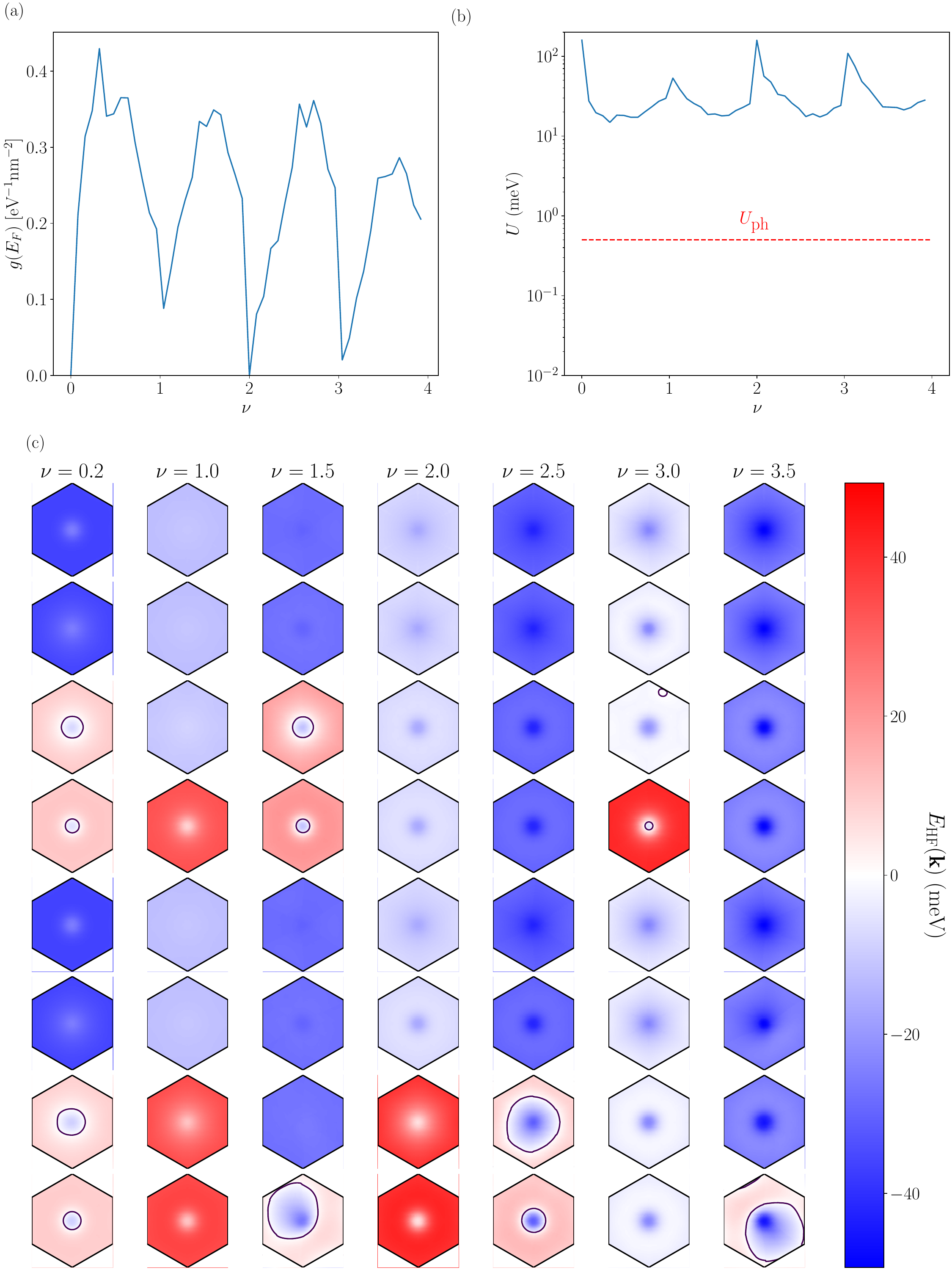}
    \caption{\textbf{Bandstructure (d).} (a) Density of states at the Fermi level with a temperature broadening corresponding to 2K. (b) Coulomb interaction scale $U$ compared to the phonon interaction scale $U_\textrm{ph}$. (c) Bandstructure and Fermi surface for seven representative fillings $\nu$. At each filling we plot the eight different bands (first four: up spin bands, last four: down spin bands). }
    \label{fig:SM1_3}
\end{figure}

\begin{figure}[h]
    \centering
    \includegraphics[width=0.7\textwidth]{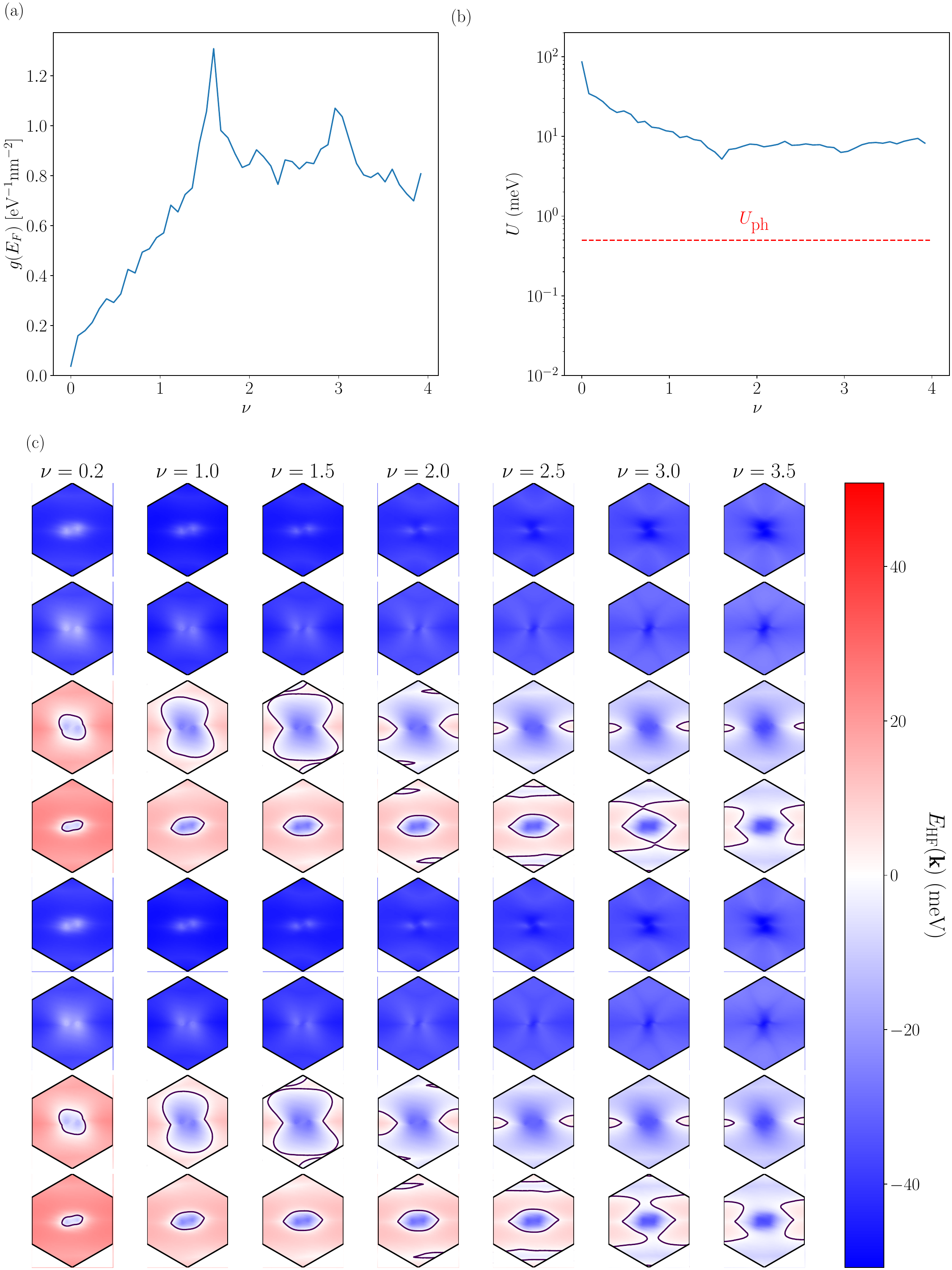}
    \caption{\textbf{Bandstructure (e).} (a) Density of states at the Fermi level with a temperature broadening corresponding to 2K. (b) Coulomb interaction scale $U$ compared to the phonon interaction scale $U_\textrm{ph}$. (c) Bandstructure and Fermi surface for seven representative fillings $\nu$. At each filling we plot the eight different bands (first four: up spin bands, last four: down spin bands). }
    \label{fig:SM1_4}
\end{figure}

\begin{figure}[h]
    \centering
    \includegraphics[width=0.7\textwidth]{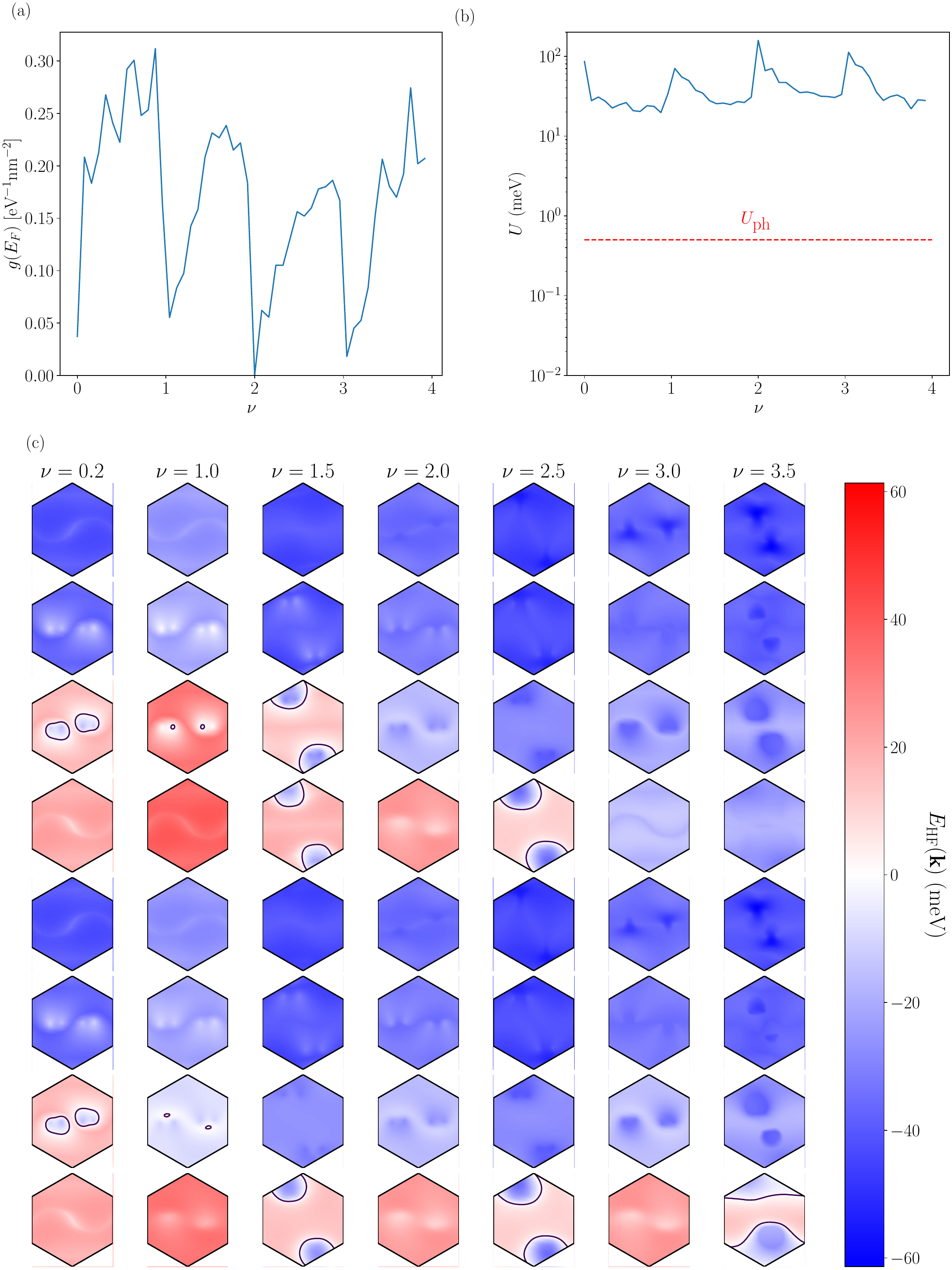}
    \caption{\textbf{Bandstructure (f).} (a) Density of states at the Fermi level with a temperature broadening corresponding to 2K. (b) Coulomb interaction scale $U$ compared to the phonon interaction scale $U_\textrm{ph}$. (c) Bandstructure and Fermi surface for seven representative fillings $\nu$. At each filling we plot the eight different bands (first four: up spin bands, last four: down spin bands). }
    \label{fig:SM1_5}
\end{figure}

For completeness we show in Figs.~\ref{fig:SM1_0}, \ref{fig:SM1_1}, \ref{fig:SM1_2}, \ref{fig:SM1_3}, \ref{fig:SM1_4}, \ref{fig:SM1_5} below the Fermi surfaces for the different bandstructures along with the density of states at the Fermi level and the Coulomb interaction scale which are reproduced from the main text (with a temperature broadening of $T=5$K.). 

\clearpage

\subsection{(a) unstrained BM model}

The unstrained BM model hosts extremely flat bands, for the parameters chosen the bandwidth is less than 2meV. This results in a very large density of states. Since we calculate the density of states with a temperature broadening of 5K, which is larger than the bandwidth, any sharp features in the density of states are washed out. This results in a gap eigenvalue $\lambda$ that does not vary significantly over the filling range (see main text). 

\clearpage

\subsection{(b) unstrained HF, $C_3$, valley $U(1)$ and spin $SU(2)$ symmetry enforced}

The Hartree-Fock state with $C_3$ symmetry enforced has the largest density of states out of all the interaction-renormalized states studied. However, this state is energetically unfavourable compared to the $C_3$ breaking state. Hartree-Fock will tend to reduce the density of states at the Fermi energy to reduce the energy, however enforcing $C_3$ symmetry forces a larger density of states, leading to a larger energy. We therefore show this state for completeness, however we expect the nematic state shown in (c) to be the more relevant parent state for superconductivity in the case where flavour symmetry breaking does not occur (e.g.~due to disorder). These bands undergo a sequence of Lifshitz transitions, where the Fermi surface topology changes.

\clearpage

\subsection{(c) unstrained HF, valley $U(1)$ and spin $SU(2)$ symmetry enforced}

There are two bands, each of which has four-fold  spin and valley degeneracy. In each spin and valley sector there are two van Hove singularities (vHSs), that are also Lifshitz transitions, in the two separate bands, where the Fermi surface goes from being simply connected, to winding around the BZ boundary. The sequence of Lifshitz transitions and van Hove singularities are reminiscent of the band structure of Bernal-stacked bilayer graphene and rhombohedral trilayer graphene with a displacement field, two other graphene-based systems that exhibit superconductivity on doping. 

\clearpage

\subsection{(d) unstrained HF}

Once flavour symmetry breaking is allowed, there are vHSs between each integer filling, where the density of states peaks. Note that this bandstructure results in the wrong Landau fan counting. Note also that the Fermi surfaces look different to those shown in \cite{IKS_PRL} due to the use a different subtraction scheme for the HF (we use the `average' subtraction scheme whereas that work used the graphene subtraction scheme). The Landau fan counting is also substraction-scheme dependent. However, independent of the subtraction scheme, the Landau fan counting at zero strain is not consistent with experiments. Including a finite amount of strain (see bandstructure (f) below) restores the consistency with experiment.

\clearpage

\subsection{(e) 0.3\% strained HF, valley $U(1)$ and spin $SU(2)$ symmetry enforced}

There are two bands, each of which has four-fold  spin and valley degeneracy. In each spin and valley sector there is a Lifshitz transition and two van Hove singularities (vHSs), that are also Lifshitz transitions, in the two separate bands, where the Fermi surface goes from being simply connected, to winding around the BZ boundary. Note that the small asymmetry between up and down spins at $\nu=3$ is a (finite-size) shell-filling effect.

\clearpage

\subsection{(f) 0.3\% strained HF}

Once again, allowing for flavour symmetry breaking leads to a sequence of vHSs between each integer filling. As shown in \cite{IKS_PRL}, there is IKS order throughout most of the range of $\nu$. The states at $\nu=2,3$ are insulating IKS states, whereas the state at $\nu=1$ is a gapless IKS state. Note that this case results in the correct Landau fan counting (4,2,1 degenerate Fermi surfaces for electron doping of fillings $\nu=0,2,3$). 

\clearpage

\section{Interaction-renormalized bandwidth}

\begin{figure}[h]
    \centering
    \includegraphics[width=0.8\textwidth]{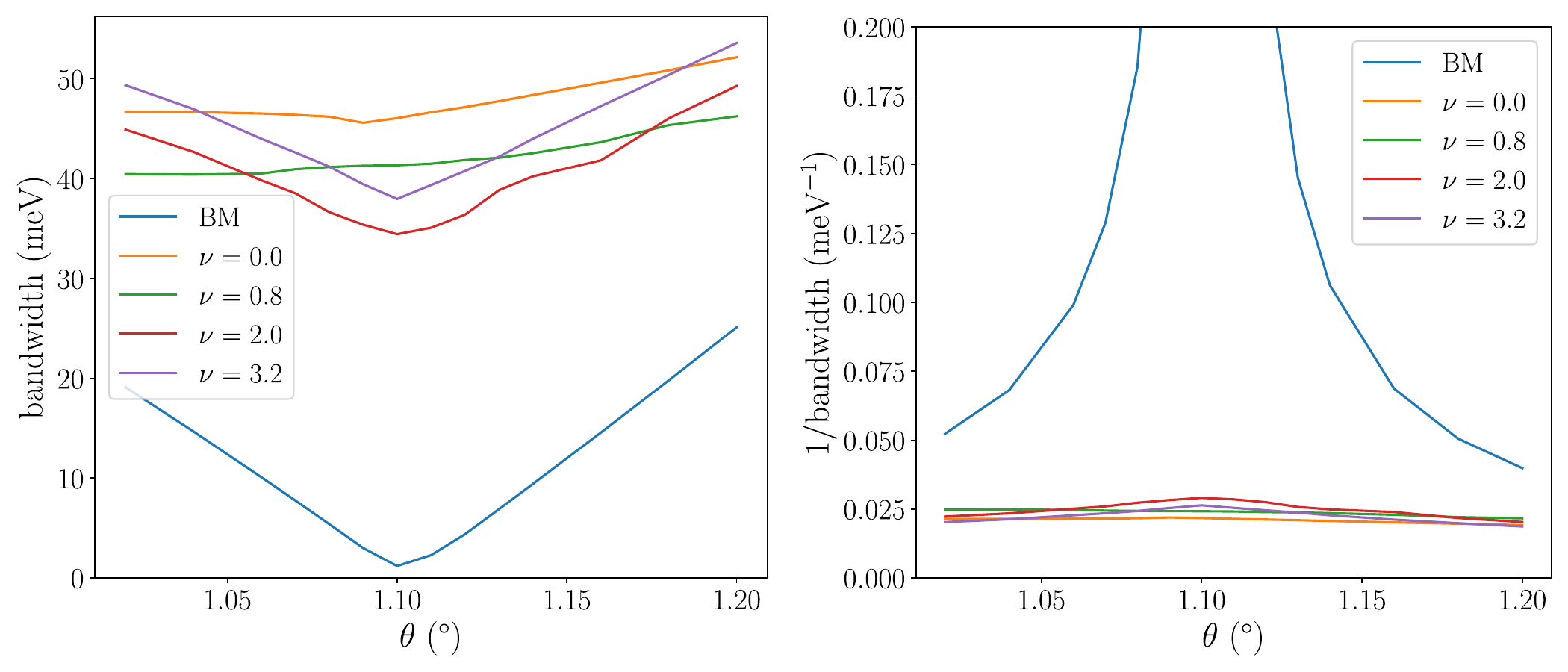}
    \caption{\textbf{Interaction renormalized bandstructure vs.~bare dispersion.} The total HF bandwidth as a function of $\theta$ for several fillings $\nu$ are shown for the unstrained model with valley-$U(1)$ symmetry enforced and zero spin polarization. Note that $C_3$ is allowed to be spontaneously broken. We also plot the inverse bandwidth, where the BM bands reach $\sim 0.8\,\text{meV}^{-1}$ (not shown on this scale).}
    \label{fig:bandwidth}
\end{figure}

In Fig.~\ref{fig:bandwidth}, we compare the interaction renormalized bandwidth with the non-interacting BM dispersion. Even at the magic angle where the BM bands have narrow bandwidth $\lesssim 1\,\text{meV}$, the interacting HF bands are significantly broadened to $\gtrsim 30\,\text{meV}$. We use the inverse of the bandwidth as a proxy for the density of states. For the BM bands, the inverse bandwidth is extremely sharply peaked about the magic angle. This would lead to a $T_c$ that is also very sharply peaked around the magic angle, whereas in experiments the twist-angle dependence is less severe \cite{Cao3}. Note that smearing due to the finite temperature may explain this broadening to some extent.

\section{General gap equation}

Consider a general Hamiltonian with momentum conservation, where we implicitly normal-order with respect to some mean-field state
\begin{gather}
    H=\sum_{k ab}h_{ab}(k)c^\dagger_{k a}c_{k b}+\frac{1}{2A}\sum_{\{a\}\{k\}}\Gamma_{abcd}(k_1,k_2,k_3)c^\dagger_{k_1 a}c^\dagger_{k_2 b}c_{k_4 d}c_{k_3 c}\\
    \Gamma_{abcd}(k_1,k_2,k_3)=\Gamma^*_{cdab}(k_3,k_4,k_1)=-\Gamma_{bacd}(k_2,k_1,k_3)=-\Gamma_{abdc}(k_1,k_2,k_4).
\end{gather}
Note that the `band' label $a$ includes flavour degrees of freedom, and the Hermiticity/antisymmetry properties of the interaction vertex $\Gamma$. If we had an unsymmetrized (but Hermitian) $V_{abcd}(k_1,k_2,k_3)$ instead then we would take
\begin{equation}
    \Gamma_{abcd}(k_1,k_2,k_3)=\frac{1}{4}\left(V_{abcd}(k_1,k_2,k_3)-V_{bacd}(k_2,k_1,k_3)-V_{abdc}(k_1,k_2,k_4)+V_{badc}(k_2,k_1,k_4)\right).
\end{equation}
Since we will eventually only need pairing momenta (zero total momentum), the relevant antisymmetrization is
\begin{equation}
    \Gamma_{abcd}(k,-k,k')=\frac{1}{4}\left(V_{abcd}(k,-k,k')-V_{bacd}(-k,k,k')-V_{abdc}(k,-k,-k')+V_{badc}(-k,k,-k')\right).
\end{equation}

Since we consider pairing at zero total momentum, we only keep interaction terms which scatter zero-momentum pairs
\begin{gather}
    H_\text{int}=\frac{1}{2A}\sum_{\{a\}kk'}U_{abcd}(k,k')c^\dagger_{k a}c^\dagger_{-k b}c_{-k' d}c_{k' c}\\
    U_{abcd}(k,k')=\Gamma_{abcd}(k,-k,k')=U^*_{cdab}(k',k)=-U_{bacd}(-k,k')=-U_{abdc}(k,-k').
\end{gather}

We define the anomalous density
\begin{equation}
    \kappa_{ab}(k)\equiv\langle c_{-k,b}c_{k,a}\rangle=-\kappa_{ba}(-k).
\end{equation}
Since $\kappa$ is antisymmetric, only a subset of its elements are independent. In particular, we could choose the independent elements to be half of the BZ (which we denote $\frac{1}{2}$BZ) but running over all band labels. For TRIM, we could choose an ordering of bands and only consider $a>b$. Later when explicitly stated, we will restrict some equations to independent elements only.

We decouple the interaction term, neglecting quadratic fluctuations
\begin{align}
    H_{\text{MF,int}}\rightarrow& -\frac{1}{2A}\sum_{\{a\}kk'}\kappa_{ab}^*(k)U_{abcd}(k,k')\kappa_{cd}(k')\\
    &+\frac{1}{2A}\sum_{\{a\}kk'} c^\dagger_{k,a}c^\dagger_{-k,b}U_{abcd}(k,k')\kappa_{cd}(k')\\
    &+\frac{1}{2A}\sum_{\{a\}kk'}c_{-k,b}c_{k,a}U_{abcd}^*(k,k')\kappa^*_{cd}(k')
\end{align}
We define the gap function
\begin{equation}
\Delta_{ab}(k)=\frac{1}{A}\sum_{k'cd}U_{abcd}(k,k')\kappa_{cd}(k')=-\Delta_{ba}(-k).
\end{equation}
Note that we could collect one momentum and two band labels into a single matrix index, and write the above as $\underline{\Delta}=\frac{1}{A}\underline{\underline{U}}\cdot\underline{\kappa}$ (so $\underline{\underline{U}}$ is Hermitian). With this, the pairing interaction becomes
\begin{equation}
    H_{\text{MF,int}}=\sum_{kab}\frac{1}{2}c^\dagger_{k,a}c^\dagger_{-k,b}\Delta_{ab}(k)+\sum_{kab}\frac{1}{2}c_{-k,b}c_{k,a}\Delta^*_{ab}(k)-\frac{1}{2}A\underline{\Delta}^*\cdot \underline{\underline{U}}^{-1}\cdot \underline{\Delta}.
\end{equation}

We now want to lift this to the path integral. Define the Nambu spinor 
\begin{equation}
    \psi_{a}(k)=[c_{k a},
    c^\dagger_{-k a}]^T.
\end{equation}
Since the existing summation in $H_\text{int}$ has redundancies, we can restrict the summation to half the BZ (we will deal with the TRIM later). Including the single-particle term, we get

\begin{equation}\label{eqn:HMF}
    H_{\text{MF}}=\sum_{k\in\frac{1}{2}\text{BZ}}\bm{\psi}^\dagger(k)\begin{pmatrix}
   h(k) & \Delta(k) \\ \Delta^\dagger(k) & -h(-k)
  \end{pmatrix}\bm{\psi}(k)-\frac{1}{2}A\underline{\Delta}^*\cdot \underline{\underline{U}}^{-1}\cdot \underline{\Delta} 
\end{equation}
\begin{equation}
   S=\int_0^\beta d\tau \left[\sum_{k\in\frac{1}{2}\text{BZ}}\bar{\bm{\psi}}(k)(\partial_\tau+\mathcal{H}(k))\bm{\psi}(k) -\frac{1}{2}A\underline{\bar{\Delta}}\cdot \underline{\underline{U}}^{-1}\cdot \underline{\Delta}\right]
\end{equation}

where $\bm{\psi}$ is a vector in band labels, and $\mathcal{H}(k)$ is the effective single-particle matrix in Nambu and band space. Note that we've lost the factor of $\frac{1}{2}$ in front of the pairing terms. In the standard way, we integrate fermion fields, assuming a static pairing field
\begin{equation}
    \mathcal{F}(\Delta,\bar{\Delta})=-T\sum_{k\in\frac{1}{2}\text{BZ},\omega_n}\text{tr}\ln \left[-i\omega_n+\mathcal{H}(k))\right] -\frac{1}{2}A\underline{\bar{\Delta}}\cdot \underline{\underline{U}}^{-1}\cdot \underline{\Delta}.
\end{equation}
where the trace acts in Nambu and band/flavor space. 

We want to minimize this free energy, so consider differentiating with respect to $\bar{\Delta}_{ab}(p)$, where $p\in\frac{1}{2}\text{BZ}$. In general, one obtains an expression that is non-linear in the gap function. Since we are interested in physics near $T_c$, we linearize with respect to the gap function. The following identities are useful: $\frac{\delta}{\delta \phi}\text{tr}\ln M^{-1}=\text{tr}\left(M\frac{\delta}{\delta\phi}M^{-1}\right)$ and the fact that the top-right component of the block matrix $\begin{pmatrix}A & B\\C & D\end{pmatrix}^{-1}$ is $-A^{-1}B(D-CA^{-1}B)^{-1}$. Working in the (mean-field) band basis, we obtain 
\begin{gather}
    \frac{\delta \text{tr}\ln}{\delta\bar{\Delta}_{ab}(p)}=-\pi_{ab}(p)\Delta_{ab}(p)\\
    \pi_{ab}(p)=T\sum_{\omega_n}\frac{1}{i\omega_n-\epsilon_a(p)}\frac{1}{-i\omega_n-\epsilon_b(-p)}=\frac{1-n_\text{F}(\epsilon_a(p))-n_\text{F}(\epsilon_b(-p))}{\epsilon_a(p)+\epsilon_b(-p)}.
\end{gather}
Note that with TRS, we recover $\pi\sim\frac{1}{2\epsilon}\tanh\frac{\epsilon}{2T}$.

Consider now the derivative of the condensation term $-\frac{1}{2}A\underline{\bar{\Delta}}\cdot \underline{\underline{U}}^{-1}\cdot \underline{\Delta}$, paying attention to momentum summations
\begin{align}
    \frac{\delta\text{cond}}{\delta\bar{\Delta}_{ab}(p)}=&-\frac{A}{2} \frac{\delta}{\delta\bar{\Delta}_{ab}(p)}\sum_{kk'cdef}\bar{\Delta}_{cd}(k)U^{-1}_{cdef}(k,k')\Delta_{ef}(k')\\
    =&-A \frac{\delta}{\delta\bar{\Delta}_{ab}(p)}\sum_{k\in\frac{1}{2}\text{BZ},k',cdef}\bar{\Delta}_{cd}(k)U^{-1}_{cdef}(k,k')\Delta_{ef}(k')\\
    =&-A\sum_{k',ef}U^{-1}_{abef}(p,k')\Delta_{ef}(k')
\end{align}
Hence we obtain the gap equation
\begin{align}
    \Delta_{ab}(k)=&-\frac{1}{A}\sum_{k',cd}U_{abcd}(k,k')\pi_{cd}(k')\Delta_{cd}(k')\\
    =&-\frac{2}{A}\sum_{k'\in\frac{1}{2}\text{BZ},cd}U_{abcd}(k,k')\pi_{cd}(k')\Delta_{cd}(k')
\end{align}
which is valid for all $k$ by antisymmetry of $\Delta$. Note that we had to be careful when un-inverting $U$ since the inverse is defined by summing over all $k$. The gap equation can be solved for just $k,k'\in\frac{1}{2}\text{BZ}$. Note that to make the kernel Hermitian, we would consider $\sqrt{\pi}\sqrt{\pi}$ on the RHS.

\section{Interaction matrix elements}
We first collect some basic facts about the BM wavefunctions
\begin{gather}
    \phi_{k\tau a}(r,I)=\frac{e^{ikr}}{\sqrt{A}}\sum_G u_{\tau a I}(k,G)e^{iGr}
\end{gather}
\begin{align}
    \psi^\dagger_{\tau,I}(r)=&e^{-i\tau Xr}\sum_{kn}\phi^*_{k\tau a}(r,I)c^\dagger_{k\tau a}\\
    =&e^{-i\tau Xr}\frac{1}{\sqrt{A}}e^{-iGr}\sum_{ka}e^{-ikr}u^*_{\tau a I}(k,G)c^\dagger_{k\tau a}
\end{align}
\begin{equation}
    u(\bm{k}+\bm{G}',\bm{G})=u(\bm{k},\bm{G}+\bm{G}').
\end{equation}
where 
$c^\dagger_{k\tau a}$ is a BM operator for band $a$, $I=(l,\sigma)$, $A$ is the total system area, and the spin $s$ is implicit.
\subsection{Coulomb part}
Let $V(q)$ be the interaction potential (which may be screened). Then the (intravalley) Coulomb Hamiltonian is
\begin{align}
    H_\text{C}=&\frac{1}{2A}\sum_{q\in\text{all}}\sum_{k,k'}\sum_{ff'\{a\}}V(q)\lambda_{\tau,ab}(k,q)\lambda^*_{\tau',dc}(k',q)\\
    &\quad\quad\times c^\dagger_{fa}(k)c^\dagger_{f'c}(k'+q)c_{f'd}(k')c_{fb}(k+q)
\end{align}
\begin{gather}
\lambda_{\tau,ab}(\bk,\bq)=\sum_{\bG I}u^*_{\tau aI}(\bk,\bG)u_{\tau bI}(\bk+\bq,\bG)=\lambda_{\tau,ba}^*(\bk+\bq,-\bq)
\end{gather}
where $f=(\tau,s)$. Recall that $q$ sums over all momenta consistent with the plane wave cutoff.

\subsection{$A_1$ phonon part}
Consider the phonon-induced electron-electron interaction due to $A_1$ phonons at momentum $K_\text{D}$
\begin{align}
    H_\text{ph}=&-g\sum_l\int d^2r\left[\left(\psi^\dagger_l\tau_x\sigma_x\psi_l\right)^2
    +\left(\psi^\dagger_l\tau_y\sigma_x\psi_l\right)^2\right]\\
    =&-4g\sum_l\int d^2r\left(\sum_s\psi^\dagger_{\bar{K}sl}\sigma_x\psi_{Ksl}\right)\left(\sum_{s'}\psi^\dagger_{Ks'l}\sigma_x\psi_{\bar{K}s'l}\right)
\end{align}
where in the final line above, $\psi$ is a spinor is sublattice space. $g$ is the coupling constant $\simeq 69 \,\text{meVnm}^2$. The flavour/species structure can be understood by remembering that the phonons are restricted to each layer. Expanding in BM eigenstates, we obtain
\begin{align}
    H_\text{ph}=&-\frac{4g}{A}\sum_{q\in\text{all}}\sum_{k,k'}\sum_{ss'\{a\}}\left[\sum_l\omega_{K,ab,l}(k,q)\omega^*_{K,dc,l}(k',q)\right]\\
    &\quad\quad\times c^\dagger_{Ksa}(k)c^\dagger_{\bar{K}s'c}(k'+q)c_{Ks'd}(k')c_{\bar{K}sb}(k+q)\\
    =&-\frac{2g}{A}\sum_{q\in\text{all}}\sum_{k,k'}\sum_{\tau ss'\{a\}}\left[\sum_l\omega_{\tau,ab,l}(k,q)\omega^*_{\tau,dc,l}(k',q)\right]\\
    &\quad\quad\times c^\dagger_{\tau sa}(k)c^\dagger_{\bar{\tau}s'c}(k'+q)c_{\tau s'd}(k')c_{\bar{\tau}sb}(k+q)
\end{align}
\begin{gather}
\omega_{\tau,ab,l}(\bk,\bq)=\sum_{\bG \sigma\sigma'}u^*_{\tau a, l\sigma}(\bk,\bG)(\sigma_x)_{\sigma,\sigma'}u_{\bar{\tau} b,l\sigma'}(\bk+\bq,\bG)=\omega_{\bar{\tau},ba,l}^*(\bk+\bq,-\bq).
\end{gather}
So $\omega$ is the intervalley analog of the usual intravalley form factor, but with uncontracted layer $l$, and sublattice contraction with $\sigma_x$. 

\subsection{Hartree-Fock basis}
In HF, we perform calculations with a boost vector $Q$, which means that we hybridize states with momentum $k$ in valley $K$ with states $k+Q$ in valley $\bar{K}$. Since we want to pair time-reversal related states, we consider a tilded momentum $\tilde{k}$ such that the lab-frame momentum (with respect to the original $\Gamma_\text{M}$) is 
\begin{equation}
    k=\tilde{k}-\tau \frac{Q}{2}\equiv \tk-b_\tau.
\end{equation} 
HF gives as outputs the following HF orbitals
\begin{align}
    \tilde{P}_{fa;f'a'}(\tilde{\bk})=\sum_{\alpha\in\text{occ}}v^*_{\alpha;fa}(\tilde{\bk})v_{\alpha;f'a'}(\tilde{\bk})\\
    d^\dagger_{n}(\tk)=\sum_{f a}v_{\alpha;fa}(\tk) c^\dagger_{fa}(\tk-b_\tau)\\
    c^\dagger_{fa}(k)=\sum_{\alpha}v^*_{\alpha;fa}(k+b_\tau)d^\dagger_\alpha(k+b_\tau)
\end{align}
where $\alpha$ indexes the HF bands at a given $\tilde{k}$. Note that we don't need to explicitly worry about the boosting if the Bloch coefficients have already been boosted by $b_\tau$. From now on, we assume that the all quantities have already been properly boosted, and drop the tildes.

\subsubsection{Coulomb part}
The Coulomb Hamiltonian in the HF basis is
\begin{align}
    H_\text{C}=&\frac{1}{2A}\sum_{q\in\text{all}}\sum_{k,k'}\sum_{ff'\{a\}}V(q)\lambda_{\tau,ab}(k,q)\lambda^*_{\tau',dc}(k',q)\\
    &\quad\times \sum_{\{\alpha\}}v^*_{\alpha;fa}(k)v^*_{\gamma;f'c}(k'+q)v_{\delta;f'd}(k')v_{\beta;fb}(k+q)\\
    &\quad\times d^\dagger_{\alpha}(k)d^\dagger_{\gamma}(k'+q)d_{\delta}(k')d_{\beta}(k+q).
\end{align}
A more concise rewriting is in terms of the HF Bloch coefficients and form factors
\begin{gather}
    w_{\alpha,fI}(k,G)=\sum_{a}v_{\alpha;fa}(k)u_{\tau a, I}(k,G)\\
    \Lambda_{\alpha\beta}(k,q)=\sum_{\bG fI}w^*_{\alpha,fI}(\bk,\bG)w_{\beta,fI}(\bk+\bq,\bG)=\Lambda^*_{\beta\alpha}(k+q,-q)
\end{gather}
leading to
\begin{align}
    H_\text{C}=&\frac{1}{2A}\sum_{k,k',q}\sum_{\{\alpha\}}\left[\sum_{G}V(q,G)\Lambda_{\alpha\beta}(k,q,G)\Lambda^*_{\delta\gamma}(k',q,G)\right]\\
    &\quad\times d^\dagger_{\alpha}(k)d^\dagger_{\gamma}(k'+q)d_{\delta}(k')d_{\beta}(k+q).
\end{align}
The interaction vertex is manifestly Hermitian.

\subsubsection{Phonon part}
The phonon-induced Hamiltonian in the HF basis is
\begin{align}
    H_\text{ph}
    =&-\frac{2g}{A}\sum_{q\in\text{all}}\sum_{k,k'}\sum_{\tau ss'\{a\}}\left[\sum_l\omega_{\tau,ab,l}(k,q)\omega^*_{\tau,dc,l}(k',q)\right]\\
    &\quad\times \sum_{\{\alpha\}}v^*_{\alpha;\tau sa}(k)v^*_{\gamma;\bar{\tau}s'c}(k'+q)v_{\delta;\tau s'd}(k')v_{\beta;\bar{\tau}sb}(k+q)\\
    &\quad\quad\times d^\dagger_{\alpha}(k)d^\dagger_{\gamma}(k'+q)d_{\delta}(k')d_{\beta}(k+q).
\end{align}
We define the HF intervalley form factor
\begin{gather}
    \Omega_{\tau l,\alpha\beta}(k,q)=\sum_{\bG s\sigma\sigma'}w^*_{\alpha,\tau s l\sigma}(\bk,\bG)(\sigma_x)_{\sigma,\sigma'}w_{\beta,\bar{\tau} s l\sigma'}(\bk+\bq,\bG)=\Omega^*_{\bar{\tau} l,\beta\alpha}(k+q,-q)
\end{gather}
leading to the concise form
\begin{align}
    H_\text{ph}=&-\frac{2g}{A}\sum_{k,k',q}\sum_{\{\alpha\}}\left[\sum_{G\tau l}\Omega_{\tau l,\alpha\beta}(k,q,G)\Omega^*_{\tau l,\delta\gamma}(k',q,G)\right]\\
    &\quad\times d^\dagger_{\alpha}(k)d^\dagger_{\gamma}(k'+q)d_{\delta}(k')d_{\beta}(k+q).
\end{align}
The interaction vertex is manifestly Hermitian.

\subsection{Pairing vertex}
We now write down the relevant matrix elements for the pairing interaction. We first summarize the unsymmetrized (but Hermitian) interaction $V$
\begin{align}
    V_{\alpha\beta\gamma\delta}(k_1,k_2,k_3)=&\left[\sum_{G}V(k_3-k_1,G)\Lambda_{\alpha\gamma}(k_1,k_3-k_1,G)\Lambda^*_{\delta\beta}(k_1+k_2-k_3,k_3-k_1,G)\right]\\
    -4g&\left[\sum_{G\tau l}\Omega_{\tau l,\alpha\gamma}(k_1,k_3-k_1,G)\Omega^*_{\tau l,\delta\beta}(k_1+k_2-k_3,k_3-k_1,G)\right].
\end{align}
We only need $k=k_1=-k_2$ and $k'=k_3=-k_4$ for pairing
\begin{align}
    V_{\alpha\beta\gamma\delta}(k,-k,k')=&\left[\sum_{G}V(k'-k,G)\Lambda_{\alpha\gamma}(k,k'-k,G)\Lambda^*_{\delta\beta}(-k',k'-k,G)\right]\\
    -4g&\left[\sum_{G\tau l}\Omega_{\tau l,\alpha\gamma}(k,k'-k,G)\Omega^*_{\tau l,\delta\beta}(-k',k'-k,G)\right].
\end{align}

From this we construct the symmetrized pairing interaction
\begin{equation}
    U_{\alpha\beta\gamma\delta}(k,k')=\frac{1}{4}\left(V_{\alpha\beta\gamma\delta}(k,-k,k')-V_{\beta\alpha\gamma\delta}(-k,k,k')-V_{\alpha\beta\delta\gamma}(k,-k,-k')+V_{\beta\alpha\delta\gamma}(-k,k,-k')\right).
\end{equation}

Let's summarize the relevant intermediate equations once and for all, recalling that all momenta entering the Bloch coefficients have been appropriately boosted to the tilded frame
\begin{align}
    V_{\alpha\beta\gamma\delta}(k,-k,k')=&\bigg[\sum_{G}V(k'-k,G)\Lambda_{\alpha\gamma}(k,k'-k,G)\Lambda^*_{\delta\beta}(-k',k'-k,G)\\
    &-4g\sum_{G\tau l}\Omega_{\tau l,\alpha\gamma}(k,k'-k,G)\Omega^*_{\tau l,\delta\beta}(-k',k'-k,G)\bigg]
\end{align}
\begin{gather}
d^\dagger_{n}(k)=\sum_{f a}v_{\alpha;fa}(k) c^\dagger_{fa}(k)\quad \text{(HF orbitals)}\\
    w_{\alpha,fI}(k,G)=\sum_{a}v_{\alpha;fa}(k)u_{\tau a, I}(k,G)\quad \text{(HF Bloch coefficients)}\\
    \Lambda_{\alpha\beta}(k,q)=\sum_{\bG fI}w^*_{\alpha,fI}(\bk,\bG)w_{\beta,fI}(\bk+\bq,\bG)=\Lambda^*_{\beta\alpha}(k+q,-q)\quad\text{(Coulomb HF form factors)}\\
    \Omega_{\tau l,\alpha\beta}(k,q)=\sum_{\bG s\sigma\sigma'}w^*_{\alpha,\tau s l\sigma}(\bk,\bG)(\sigma_x)_{\sigma,\sigma'}w_{\beta,\bar{\tau} s l\sigma'}(\bk+\bq,\bG)=\Omega^*_{\bar{\tau} l,\beta\alpha}(k+q,-q)\quad\text{(phonon HF form factors)}.
\end{gather}
Note that $q$ in the form factors above have both an mBZ and a RLV part.

\section{Thomas Fermi-screening}
The RPA-screened interaction takes the form
\begin{equation}
    V(q)=\frac{V^0(q)}{1-\Pi(q)V^0(q)}
\end{equation}
where the bare (gate-screened) Coulomb interaction is 
\begin{equation}
    V^0(q)=\frac{e^2}{2\epsilon q}\tanh(qd_{sc})
\end{equation}
In the Thomas-Fermi approximation we take the polarization function to be a constant 
\begin{equation}
    \Pi(q)\approx -D(E_F)=\frac{\epsilon}{e^2q_{TF}}.
\end{equation}

\section{Hartree-Fock interpolation}
We perform self-consistent HF on a $10\times10$ momentum grid, however solving the gap equation for temperatures $T\lesssim5K$ requires the bandstructure on a finer grid. To this end we employ the interpolation scheme used in Ref.~\cite{TSTG2}. The interaction Hamiltonian is
\begin{align}
    \hat{H}_{\text{int}}&=\frac{1}{2}
    \sum_{\stackrel{ss'\tau\tau'}{abcd}}
    \sum_{\bm{k}^\alpha\bm{k}^\beta\bm{q}\bm{G}}
    \lambda_{\tau,a,b}(\bm{k}^\alpha;\bm{q},\bm{G})
    M_{\tau',d,c}(\bm{k}^\beta;\bm{q},\bm{G})\\
    &\quad\quad\quad\quad\quad\quad\quad  \times 
    c^\dagger_{\tau as}(\bm{k}^\alpha)
    c^\dagger_{\tau' cs'}(\bm{k}^\beta+\bm{q})
    c_{\tau' ds'}(\bm{k}^\beta)
    c_{\tau bs}(\bm{k}^\alpha+\bm{q}).
    \\
    M_{\tau,a,b}(\bm{k};\bm{q},\bm{G})
    &\equiv\lambda^*_{\tau,a,b}(\bm{k};\bm{q},\bm{G})V(\bm{q},\bm{G})
\end{align}
and where the form factors are
\begin{equation}
\lambda_{\tau,a,b}(\bm{k};\bm{q},\bm{G})=\braket{u_{\tau,a}(\bm{k})|u_{\tau,b}(\bm{k}+\bm{q}+\bm{G})}
\end{equation}
Assuming translational invariance (in the boosted frame), we have to set $\bm{q}=0$ in the Hartree part of the Hamiltonian which becomes
\begin{align}
    \hat{H}^{HF,dir}&=\sum_{\stackrel{ss'\tau\tau'}{abcd}}\sum_{\bm{k}^\alpha\bm{k}^\beta \bm{G}}
    \lambda_{\tau,a,b}(\bm{k}^\alpha;0,\bm{G})
    M_{\tau',d,c}(\bm{k}^\beta;0,\bm{G})
    P_{\tau' cs';\tau' ds'}(\bm{k}^\beta)\\
    &\quad\quad\quad\quad\quad\quad\quad
    \times c^\dagger_{\tau as}(\bm{k}^\alpha)
    c_{\tau bs}(\bm{k}^\alpha)  
\end{align}
In the Fock part we have to set $\bm{k}^\alpha=\bm{k}^\beta\equiv\bm{k}$ such that it becomes
\begin{align}
    \hat{H}^{HF,ex}&=\sum_{\stackrel{ss'\tau\tau'}{abcd}}\sum_{\bm{k} \bm{q}\bm{G}}
    \lambda_{\tau,a,b}(\bm{k};\bm{q},\bm{G})
    M_{\tau',d,c}(\bm{k};\bm{q},\bm{G})
    P_{\tau' cs'; \tau bs}(\bm{k}+\bm{q})\\
    &\quad\quad\quad\quad\quad\quad\quad
    \times c^\dagger_{\tau as}(\bm{k})
    c_{\tau' ds'}(\bm{k})  
\end{align}
Assuming colinear spins, we can rewrite this as 
\begin{equation}
    \hat{H}^{HF}=\sum_{s\tau\tau'ad\bm{k}}(h^{dir}_{s;\tau a,\tau'd}(\bm{k})-h^{ex}_{s;\tau a,\tau'd}(\bm{k}))c^\dagger_{\tau as}(\bm{k})
    c_{\tau' ds}(\bm{k})  
\end{equation}
where
\begin{equation}
    h^{dir}_{s;\tau a,\tau''b}(\bm{k})=\delta_{\tau\tau''}\sum_{\stackrel{s'\tau'}{cd}}\sum_{\bm{k}^\beta \bm{G}}V(0,\bm{G})
    \lambda_{\tau,a,b}(\bm{k};0,\bm{G})
    \lambda^*_{\tau',d,c}(\bm{k}^\beta;0,\bm{G})
    P_{s';\tau' c,\tau' d}(\bm{k}^\beta)
\end{equation}
and 
\begin{equation}
    h^{ex}_{s;\tau a,\tau'd}(\bm{k})=\sum_{bc}\sum_{\bm{q}\bm{G}}
    V(\bm{q},\bm{G})\lambda_{\tau,a,b}(\bm{k};\bm{q},\bm{G})
    \lambda^*_{\tau',d,c}(\bm{k};\bm{q},\bm{G})
    P_{s;\tau' c,\tau b}(\bm{k}+\bm{q})
\end{equation}

The idea of the interpolation scheme is that the momentum argument of the projector ($\bm{k}^\beta$ and $\bm{k+q}$ respectively) lies on a coarse grid, while the momentum $\bm{k}$ can lie on a much finer mesh. 

\section{Importance sampling of the gap equation}

In Fig. 1 of the main text, we evaluate the eigenvalue $\lambda$ of the gap equation at $T=5K$, which corresponds to an energy $k_BT=0.4$meV. For the dispersive bandstructures (b-f), the region in momentum space where the states lie within $\sim k_BT$ of the Fermi level is therefore small. In order to have enough points close to the Fermi level, we therefore have to employ the interpolation scheme described above. We interpolate the bandstructure to a $40\times 40$ grid and solve the gap equation on that grid. However, in the absence of additional symmetries, the solution to the gap equation 
\begin{equation}
    \Delta_{ab}(k)=-\frac{1}{A}\sum_{k',cd}U_{abcd}(k,k')\pi_{cd}(k')\Delta_{cd}(k')
\end{equation}
becomes computationally expensive when including all momentum points and all bands. Using an energy cutoff $E_c$ such that only points within $E_c$ of the Fermi energy are included is not a suitable option since the polarization function (which dictates how sharply peaked the summand is around the Fermi surface) decays algebraically rather than exponentially. In order to reduce the computational cost by taking advantage of the fact that the polarization function is sharply peaked around the Fermi surface, we evaluate the sum over $\vec{k}'$ in the gap equation via Monte-Carlo sampling. We use the normalized polarization function  
\begin{equation}
    p(\vec{k})=\frac{\sum_{ab}\pi_{ab}(\vec{k})}{\sum_{ab\vec{k}}\pi_{ab}(\vec{k})}
\end{equation}
as the probability distribution for the importance sampling. On the other hand, for the BM model bandstructure (a), the bandwidth is significantly smaller, such that electrons throughout the Brillouin zone (BZ) can contribute to pairing. In that case we therefore include all momentum points in the BZ when solving the gap equation, but we solve the gap equation on a $10\times 10$ grid instead of a $40\times40$ grid.

\section{Symmetries of the gap function}
In the case where we impose the flavour symmetries, we have spin $SU(2)$ and valley $U(1)$ symmetry. Therefore, the gap function i.e.~the eigenvector of the gap equation can be labelled by the $S_z$ and $S^2$ eigenvalue as well as the $\tau_z$ eigenvalue. We consider only $\tau_z=0$ pairing. The valley singlet and valley triplet can mix. Inversion acts as $\tau_x$ and takes $\vec{k}\to-\vec{k}$. In order to satisfy the correct fermionic exchange statistics, the spin singlet must be even parity. This can be either the valley singlet combined with a function that is odd in momentum in the moir\'e Brillouin zone (mBZ) or the valley triplet combined with a function that is even in momentum in the mBZ. The spin triplet must be odd parity. This can be the valley singlet combined with an mBZ-even function or the valley triplet combined with the mBZ-odd function. Thus, the eigenfunctions take the following form:
\begin{enumerate}
    \item spin singlet ($S^2=0$, $S_z=0$): $\Delta\sim\frac{\ket{\uparrow\downarrow}-\ket{\downarrow\uparrow}}{\sqrt{2}}\bigg[\frac{\ket{K\bar K}-\ket{\bar K K}}{\sqrt{2}}\Delta_o(\vec{k})+\frac{\ket{K\bar K}+\ket{\bar K K}}{\sqrt{2}}\Delta_e(\vec{k})\bigg]$
    \item spin triplet ($S^2=1$, $S_z=0$): $\Delta\sim\frac{\ket{\uparrow\downarrow}+\ket{\downarrow\uparrow}}{\sqrt{2}}\bigg[\frac{\ket{K\bar K}-\ket{\bar K K}}{\sqrt{2}}\Delta_e(\vec{k})+\frac{\ket{K\bar K}+\ket{\bar K K}}{\sqrt{2}}\Delta_o(\vec{k})\bigg]$
    \item spin triplet ($S^2=1$, $S_z=1$): $\Delta\sim\ket{\uparrow\uparrow}\bigg[\frac{\ket{K\bar K}-\ket{\bar K K}}{\sqrt{2}}\Delta_e(\vec{k})+\frac{\ket{K\bar K}+\ket{\bar K K}}{\sqrt{2}}\Delta_o(\vec{k})\bigg]$
    \item spin triplet ($S^2=1$, $S_z=-1$): $\Delta\sim\ket{\downarrow\downarrow}\bigg[\frac{\ket{K\bar K}-\ket{\bar K K}}{\sqrt{2}}\Delta_e(\vec{k})+\frac{\ket{K\bar K}+\ket{\bar K K}}{\sqrt{2}}\Delta_o(\vec{k})\bigg]$
\end{enumerate}
where $\Delta_e(\vec{k})=\Delta_e(-\vec{k})$ is an even parity function in the mBZ, while $\Delta_o(\vec{k})=-\Delta_o(-\vec{k})$ is an odd parity function in the mBZ. The spin triplet eigenfunctions are all degenerate due to the spin $SU(2)$ symmetry. The spin triplet and spin singlet eigenfunctions are degenerate due to the combined spin-valley $SU(2)_K\times SU(2)_{\bar K}$.

\section{Relationship to Coulomb-driven band flattening}

In Refs.~\cite{Lewandowski2021does,Choi} a mechanism for Coulomb-driven band flattening has been proposed. In that case, the non-interacting bandstructure is chosen to have a relatively large dispersion of $\gtrsim20$meV by tuning away from the flat-band limit (e.g.~by varying the twist angle and the chiral ratio). Upon doping electrons into the conduction band, the Hartree (and Fock) corrections lead to a dip at the $\Gamma$ point in the mBZ. This dip  competes with the BM model peak in the conducting band at $\Gamma$. These two competing effects lead to a flattening of the bands. For this effect to manifest, one requires the non-interacting dispersion to be on the order of the interaction scale. However, we note that even with this band-flattening effect, the bandwidth stays above $\sim10$meV, and is therefore significantly larger than the bands used for the superconductivity calculations in \cite{Wu,liu2023electronkphonon}. In our case we start off with BM model parameters yielding a very small bandwidth similar to the bandstructures used in \cite{Wu,liu2023electronkphonon}. In this case, the interaction energy scale is significantly larger than the bandwidth and the interactions always have a band-unflattening effect. 

Compared to our work where we perform a BCS calculation of pairing from optical phonons, \cite{Lewandowski2021does} performs an Eliashberg calculation of acoustic phonon-mediated pairing. However \cite{Lewandowski2021does} finds a maximum critical temperature around 0.3K, which is an order of magnitude lower than the highest $T_c$ reported in experiments which is around 3K \cite{Cao3}. The strong Coulomb repulsion will significantly reduce the critical temperature. In fact, disorder and strain are both likely significant in TBG and would reduce the critical temperature from this mechanism even further. This is in line with our conclusion that the interaction renormalized bandwidth is too large to account for the high $T_c$ seen in experiments.

\end{appendix}
\end{document}